\documentstyle[12pt,epsf]{article}

\def\Frac#1#2{\frac{\displaystyle{#1}}{\displaystyle{#2}}}
\def\lsim{\raise0.3ex\hbox{$\;<$\kern-0.75em\raise-1.1ex\hbox{$\sim\;$}}}
\def\gsim{\raise0.3ex\hbox{$\;>$\kern-0.75em\raise-1.1ex\hbox{$\sim\;$}}}

\def\npb#1#2#3{    {\it Nucl. Phys. }{\bf B #1} (19#2) #3}
\def\plb#1#2#3{    {\it Phys. Lett. }{\bf B #1} (19#2) #3}
\def\prd#1#2#3{    {\it Phys. Rev. }{\bf D #1} (19#2) #3}
\def\prdII#1#2#3{    {\it Phys. Rev. }{\bf D #1} (20#2) #3}
\def\prep#1#2#3{   {\it Phys. Rep. }{\bf #1} (19#2) #3}
\def\prl#1#2#3{    {\it Phys. Rev. Lett. }{\bf #1} (19#2) #3}

\def\zpc#1#2#3{    {\it Zeit. f\"ur Physik }{\bf C #1} (19#2) #3}

\def\ibid#1#2#3{   {\it ibid. }{\bf #1} (19#2) #3}
\newcommand\jhep[3]  {{\it J. High Energy Phys.\ }{\bf #1} (#2) #3}

\def\Im{\mathop{\mbox{Im}}}

\newcommand{\beqn}{\begin{eqnarray}}
\newcommand{\beq}{\begin{equation}}
\newcommand{\eeqn}{\end{eqnarray}}
\newcommand{\eeq}{\end{equation}}
\newcommand{\nn}{\nonumber}

\begin{document}
\hyphenation{re-pha-sings bo-unds e-lec-tron u-ni-ver-sa-li-ty
sfer-mions in-de-pen-dent Yu-ka-wa se-arch char-gi-no
spec-ci-fically su-per-sym-me-tric pha-ses model-in-de-pen-dent diago-na-lity 
                   an-ni-hi-la-tion pen-guin}
\begin{titlepage}
\begin{flushright}
SISSA/20/2000/EP
\end{flushright}
\vskip 1.5cm
\centerline {\Large{\bf New Physics behind the }}
\vskip 0.2cm
\centerline {\Large{\bf Standard Model's door?$^{*}$}} \normalsize
 
\vskip 1cm
\centerline {A. Masiero and O. Vives}
\vskip 0.2cm
\centerline {\it SISSA, Via Beirut 2--4, 34013 Trieste, Italy and}
\vskip 0.2cm
\centerline {\it INFN, sez. di Trieste, Trieste, Italy}

\vskip 1cm
 
\begin{abstract}
We review the main reasons pushing us beyond the SM and we argue in favor of 
new physics at the electroweak scale (hence experimentally accessible at 
present or near--future machines). We focus on the appealing possibility 
that such new physics is given by a supersymmetric (SUSY) extension of the SM. 
We discuss the minimal case, Constrained Minimal Supersymmetric SM, and more 
general (maybe more natural) 
cases where some of the drastic assumptions of the CMSSM are dropped. 
In particular, in these lectures we focus on CP violation and its relation 
to flavor physics in the SUSY context. CP constrains the low--energy SUSY 
extensions of the SM, but, at the same time, it provides new powerful tool 
for indirect SUSY searches.
\end{abstract}
\vskip 1cm
\centerline{\small{$^{*}$ Lectures given by A. Masiero at the International
  School on Subnuclear Physics, }} 
\centerline{\small{37th Course: ``Basics and Highlights in Fundamental 
Physics'',}}
\centerline{\small{Erice, Italy, 29 August--7 September 1999}} 

\vfill

\end{titlepage}

\newpage
\section{Introduction}
\label{sec:intro}

The success of the standard model (SM) predictions is remarkably 
high and, indeed, to some extent, even beyond what we theorists would
have expected. A common view before LEP started operating was that some
new physics related to the electroweak symmetry breaking should show up
when precisions at the percent level on some electroweak observable could
be reached. As we know, on the contrary, even reaching sensitivities 
better than the percent has not given rise to any indication of departure
from the SM predictions. All that can be summarized in a powerful
statement about the ``low--energy'' limit of any kind of new physics beyond
the SM: such new physics has to reproduce the SM with great accuracy when
we consider its limit at energy scales of the order of the electroweak
scale. 

The fact that with the SM we have a knowledge of fundamental interactions
up to energies of ${\cal{O}}(100)$ GeV should not be underestimated: it represents
a tremendous and astonishing success of our gauge theory approach in
particle physics and it is clear that it represents one of the great
achievements in a century of great conquests in physics. Having said that,
we are now confronting ourselves with an embarrassing question: if the SM
is so extraordinarily good, does it make sense do go beyond it? The
answer, in our view, is certainly positive. This ``yes'' is not only
motivated by what we could define ``philosophical'' reasons, but there are
specific motivations pushing us beyond the SM: we will group them in two
broad categories, theoretical and ``observational'' reasons.   

\subsection{Theoretical reasons for new physics}
\label{sec:theore}

The major theoretical conundra of the SM are related the following issues:
flavor problem, unification of the fundamental interactions and gauge
hierarchy problem. We briefly remind what they are about.

\vskip .65cm
{\bf Flavor problem}. All the masses and mixings of fermions are just free
(unpredicted) parameters in the SM. To be sure, there is not even any
hint in the SM about the number and rationale of fermion families. Leaving
aside predictions for individual masses, not even a rough relation
among fermion masses within the same generation or among different
generations is present.

\vskip .65cm
{\bf Unification of forces}. At the time of the Fermi theory we had two
couplings to describe the electromagnetic and the weak interactions (the
electric constant and the Fermi constant, respectively). In the SM we are
trading off those two couplings with two new couplings, the gauge
couplings of  $SU(2)$ and $U(1)$. Moreover, the gauge coupling of the
strong interactions is very different from the other two. We cannot say
that the SM represents a true unification of fundamental interactions,
even leaving aside the problem that gravity is not considered at all by
the model.

\vskip .65cm
{\bf Gauge hierarchy}. Fermion and vector boson masses are ``protected'' by
symmetries in the SM (i.e., their masses can arise only when we break certain
symmetries). On the contrary, the Higgs scalar mass does not enjoy such a
symmetry protection. We would expect such mass to naturally jump to some
higher scale where new physics sets in (this new energy scale could be
some grand unification scale or the Planck mass, for instance). The only
way to keep the Higgs mass at the electroweak scale is to perform
incredibly accurate fine tunings of the parameters of the scalar sector. 

\vskip .8cm   
\subsection{``Observational'' reasons for new physics}
\label{sec:observ}
We have already said that all the experimental particle physics results of
these last years have marked one success after the other of the SM. What
do we mean then by ``observational'' difficulties for the SM? It is
curious that such difficulties do not arise from observations within the
strict particle physics domain, but rather they originate from possible  
{\bf ``clashes''} of the particle physics SM with the standard model of 
cosmology (i.e., the Hot Big Bang) or the standard model of the Sun. 
Explicitly we have in mind the following points.

\vskip .65cm
{\bf Dark Matter}. Denoting with $\Omega$ the ratio of the energy density
to the critical energy density, the problem of dark matter (DM) can be
summarized in the following two numbers: $\Omega_{DM} = 0.3$ and $\Omega_B
< 0.1$. The first number denotes the amount of the contribution to
$\Omega$ due to DM as inferred from measurements at the level of cluster
of galaxies. The upper bound denotes the highest contribution of baryonic
matter to $\Omega$ to have compatibility with one of the main pillars of
the Big Bang model: nucleosynthesis. The clash between the above two
numbers underlines the fact that we definitely need a large amount of
non--baryonic DM. In the electroweak SM, no viable non--baryonic candidate
exists to fulfill this task (remember that in the SM neutrinos are
strictly massless). Hence, the existence of a (large) amount of
non--baryonic DM pushes us to introduce new particles in addition to those of
the SM.
\vskip .65cm
{\bf Baryogenesis}. Given that we have strong evidence that the Universe
is vastly matter--antimatter asymmetric (i.e. no sizeable amount of
primordial antimatter has survived), it is appealing to have a dynamical
mechanism to give rise to such large baryon--antibaryon asymmetry starting
from a symmetric situation. In the SM it is not possible to have such an
efficient mechanism for baryogenesis. In spite of the fact that at the
quantum level sphaleronic interactions violate baryon number in the SM,
such violation cannot lead to the observed large matter--antimatter
asymmetry (both CP violation is too tiny in the SM and, also, the present
experimental lower bounds on the Higgs mass do not allow for a
conveniently strong electroweak phase transition). Hence, a dynamical
baryogenesis calls for the presence of new particles and interactions
beyond the SM (successful mechanisms for baryogenesis in the context of
new physics beyond the SM are well known).

\vskip .65cm
{\bf Inflation}. Several serious cosmological problems (flatness,
causality, age of the Universe, ...) are beautifully solved if the early
Universe underwent some period of exponential expansion (inflation). The
SM with its Higgs doublet does not succeed to originate such an
inflationary stage. Again some extensions of the SM, where in particular
new scalar fields are introduced, are able to produce a temporary
inflation of the early Universe.       

\vskip .8cm
\subsection{The SM as an effective low--energy theory}

The above theoretical and ``observational'' arguments strongly motivate us
to go beyond the SM. On the other hand, the clear success of the SM in
reproducing all the known phenomenology up to energies of the order of the
electroweak scale is telling us that the SM has to be recovered as the
low--energy limit of such new physics.
 Indeed, it may even well be the case  that we have a ``tower''
of underlying theories which show up at different energy scales.

If we accept the above point of view, we may try to find signals of new
physics considering the SM as a truncation to renormalizable operators of
an effective low--energy theory which respects the  $SU(3)\times SU(2) \times
U(1)$
symmetry  and whose fields are just those of the SM. The renormalizable
(i.e. of canonical dimension less or equal to four) operators giving rise
to the SM enjoy three crucial properties which have no reason to be shared
by generic operators of dimension larger than four. They are the
conservation (at any order in perturbation theory) of Baryon (B) and
Lepton (L) numbers and an adequate suppression of Flavor Changing Neutral
Current (FCNC) processes through the GIM mechanism.  

Now consider the new physics (directly above the SM in the
``tower'' of new physics theories) to have a typical energy scale $\Lambda$.
In the low--energy effective Lagrangian, such scale appears with a positive
power only in the quadratic scalar term (scalar mass) and in the dimension
zero operator which can be considered a cosmological constant. Notice that
$\Lambda$ cannot appear in dimension three operators related to fermion
masses because chirality forbids direct fermion mass terms in the
Lagrangian. Then, in all operators of dimension larger than four, $\Lambda$
will show up in the denominator with powers increasing with the dimension
of the corresponding operator. 

The crucial question that all of us, theorists and experimentalists, ask
ourselves is: where is $\Lambda$? Namely is it close to the electroweak
scale (i.e. not much above $100$ GeV) or is $\Lambda$ of the order of the
grand unification scale or the Planck scale? B-- and L--violating processes
and FCNC phenomena represent a potentially interesting clue to answer this
fundamental question.

Take $\Lambda$ to be close to the electroweak scale. Then we may expect
non--renormalizable operators with B, L and flavor violations not to be
largely suppressed by the presence of powers of $\Lambda$ in the
denominator. Actually this constitutes, in general, a formidable challenge
for any model builder who wants to envisage new physics close to $M_W$.
Theories with dynamical breaking of the electroweak symmetry
(technicolour) and low--energy supersymmetry constitute examples of new
physics with a ``small'' $\Lambda$. In these lectures we will see that the
above general considerations on potentially large B, L and flavor violations
apply to the SUSY case (it is well--known that FCNC represent a major
problem also in technicolour schemes).

Alternatively, given the abovementioned potential danger of having a small
$\Lambda$, one may feel it safer to send $\Lambda$ to super--large values.
Apart from kind of ``philosophical'' objections related to the unprecedented  
gap of many orders of magnitude without any new physics, the above
discussion points out a typical problem of this approach. Since the
quadratic scalar terms have a coefficient in front scaling with
$\Lambda^2$, we expect all scalar masses to be of the order of the
super--large scale $\Lambda$. This is the gauge hierarchy problem, and it
constitutes the main (if not only) reason to believe that SUSY should be a
low--energy symmetry.

Notice that the fact that SUSY should be a fundamental symmetry of Nature
(something of which we have little doubt given the ``beauty'' of this
symmetry) 
does not imply by any means that SUSY should be a low--energy symmetry,
namely that it should hold unbroken down to the electroweak scale.
SUSY may well be present in
Nature but be broken at some very large scale (Planck scale or string
compactification scale). In that case SUSY would be of no use in tackling
the gauge hierarchy problem and its phenomenological relevance would be
practically zero. On the other hand, if we invoke SUSY to tame the growth
of the scalar mass terms with the scale $\Lambda$, then we are forced to
take the view that SUSY should hold as a good symmetry down to a scale
$\Lambda$ close to the electroweak scale. Then B, L and FCNC may be useful
for us to shed some light on the properties of the underlying theory from
which the low--energy SUSY Lagrangian resulted. Let us add that there is an
independent argument in favor of this view that SUSY should be a
low--energy symmetry. The presence of SUSY partners at low energy creates
the conditions to have a correct unification of the strong and electroweak
interactions. If they were at $M_{\rm Planck}$ and the SM were all the physics
up to super--large scales, the program of achieving such a unification
would largely fail, unless one complicates the non--SUSY GUT scheme with a
large number of Higgs representations and/or a breaking chain with
intermediate mass scales is invoked.

In the above discussion, we stressed that we are not only insisting on
the fact that SUSY should be present at some stage in Nature, but we are
asking for something much more ambitious: we are asking for SUSY to be a
low--energy symmetry, namely it should be broken at an energy scale as low
as the electroweak symmetry breaking scale. This fact can never be
overestimated. There are indeed several reasons pushing us to introduce
SUSY: it is the most general symmetry compatible with a local,
relativistic quantum field theory, it softens the degree of divergence of
the theory, it looks promising for a consistent quantum description of
gravity together with the other fundamental interactions. However, all
these reasons are not telling us where we should expect SUSY to be broken.
For that matter, we could even envisage the maybe ``natural'' possibility
that SUSY is broken at the Planck scale. What is relevant for
phenomenology is that the gauge hierarchy problem and, to some extent, the
unification of the gauge couplings are actually forcing us to ask for SUSY
to be unbroken down to the electroweak scale, hence implying that the SUSY
copy of all the known particles, the so--called s--particles should have a
mass in the $100$--$1000$ GeV mass range. If LEP and Tevatron are not going
to see any SUSY particle, at least the advent of LHC will be decisive in
establishing whether low--energy SUSY actually exists or it is just a fruit
of our (ingenious) speculations. Although even after LHC, in case of a
negative result for the search of SUSY particles, we will not be able to
``mathematically'' exclude all the points of the SUSY parameter space, we
will certainly be able to very reasonably assess whether the low--energy
SUSY proposal makes sense or not. 

Before the  LHC (and maybe Tevatron) direct searches for SUSY
signals we should ask ourselves whether we can hope to have some indirect
manifestation of SUSY through virtual effects of the SUSY particles. 

We know that, in the past, virtual effects (i.e. effects due to the
exchange of yet unseen particles in the loops) were precious in leading us
to major discoveries, like the prediction of the existence of the charm
quark or the heaviness of the top quark long before its direct
experimental observation. Here we focus on the potentialities of SUSY
virtual effects in processes which are particluraly suppressed (or
sometime even forbidden) in the SM; the flavor changing neutral current
phenomena and the processes where CP violation is violated.

\section{Flavor, CP and New Physics}
\label{sec:FCNC}

The generation of fermion masses and mixings (``flavor problem'') gives 
rise to a first and important distinction among theories of new physics 
beyond the electroweak standard model. 

One may conceive a 
kind of new physics which is completely ``flavor blind'', i.e. new 
interactions which have nothing to do with the flavor structure. To 
provide an example of such a situation, consider a scheme where flavor 
arises at a very large scale (for instance the Planck mass) while new 
physics is represented by a supersymmetric extension of the SM 
with supersymmetry broken at a much lower scale and with the SUSY 
breaking transmitted to the observable sector by flavor--blind gauge 
interactions. In this case, one may think that the new physics does not 
cause any major change to the original flavor structure of the SM, 
namely that the pattern of fermion masses and mixings is compatible with 
the numerous and demanding tests of flavor changing neutral currents.

Alternatively, one can conceive a new physics which is entangled 
with the flavor problem. As an example consider a technicolour scheme 
where fermion masses and mixings arise through the exchange of new gauge 
bosons which mix together ordinary fermions and technifermions. Here we expect 
(correctly enough) new physics to have potential problems in 
accommodating the usual fermion spectrum with the adequate suppression 
of FCNC. As another example of new physics which is not flavor blind, 
take a more conventional SUSY model which is derived from a 
spontaneously broken N=1 supergravity and where the SUSY breaking 
information is conveyed to the ordinary sector of the theory through 
gravitational interactions. In this case we may expect that the scale at 
which flavor arises and the scale of SUSY breaking are not so different 
and possibly the mechanism itself of SUSY breaking and transmission is 
flavor--dependent. Under these circumstances, we may expect 
a potential flavor problem to arise, namely that SUSY contributions to 
FCNC processes are too large.

\vskip .8cm
\subsection{The Flavor Problem in SUSY}

The potentiality of probing SUSY in FCNC phenomena was readily realized
when 
the era of SUSY  phenomenology started in the early 80's \cite{susy2}.
 In particular, the 
major implication that the scalar partners of quarks of the same electric 
charge but belonging to different generations had to share a remarkably high 
mass degeneracy was emphasized.

Throughout the large amount of work in this last decade, it became clearer 
and clearer that generically talking of the implications of low--energy SUSY 
on FCNC may be rather misleading. Even in the Minimal SUSY extension of the SM 
(MSSM) \cite{susy1} from the point of view of the particle content, we have 
a host of different situations. The so--called Constrained Minimal 
Supersymmetric Standard Model (CMSSM) is the simplest possibility, and 
the FCNC 
contributions can be computed in terms of a very limited set of unknown 
new SUSY parameters. Remarkably enough, this minimal model succeeds to
pass all  
the set of FCNC tests unscathed. To be sure, it is possible to severely 
constrain the SUSY parameter space, for instance using $b 
\to s \gamma$, in a way which is complementary to what is achieved by direct 
SUSY searches at colliders.

However, the CMSSM is by no means equivalent to low--energy SUSY. A first 
sharp distinction concerns the mechanism of SUSY breaking and 
transmission to the observable sector which is chosen. As we mentioned 
above, in models with gauge--mediated SUSY breaking (GMSB models
\cite{GMSB1,GMSB2,GMSB3})
it may be possible to avoid the 
FCNC threat ``ab initio'' (notice that this is not an automatic feature of 
this class of models, but it depends on the specific choice of the 
sector which transmits the SUSY breaking information, the so--called 
messenger sector). The other more ``canonical'' class of SUSY theories 
(including also CMSSM) has gravitational messengers and a very large 
scale at which SUSY breaking occurs. In this talk we will focus only on 
this class of gravity--mediated SUSY breaking models. Even sticking to 
this more limited choice, we have a variety of options with very 
different implications for the flavor problem. 

First, there exists an interesting large class of SUSY realizations 
where the customary R--parity (which is invoked to suppress proton decay) 
is replaced by 
other discrete symmetries which allow either baryon or lepton violating terms 
in the superpotential. But, even sticking to the more orthodox view of 
imposing R--parity, we are still left with a large variety of extensions of 
the MSSM at low energy. The point is that low--energy SUSY ``feels'' the new 
physics 
at the super--large scale at which supergravity  (i.e., local supersymmetry) 
broke down. In this last couple of years, we have witnessed an increasing 
interest in supergravity realizations without the so--called flavor 
universality of the terms which break SUSY explicitly. Another class of 
low--energy SUSY realizations, which differ from the MSSM in the FCNC sector, 
is obtained from SUSY--GUT's. The interactions involving super--heavy 
particles 
in the energy range between the GUT and the Planck scale bear important 
implications for the amount and kind of FCNC that we expect at low energy.

Even when R--parity is imposed, the FCNC challenge is not over. It is true
that in this case, analogously to what happens in the SM, no tree
level FCNC contributions arise. However, it is well--known that this is a
necessary but not sufficient condition to consider the FCNC problem
overcome. The loop contributions to FCNC in the SM exhibit the presence of
the GIM mechanism and we have to make sure that in the SUSY case with R
parity some analog of the GIM mechanism is active. 

To give a qualitative idea of what we mean by an effective super--GIM
mechanism, let us consider the following simplified situation where the
main features emerge clearly. Consider the SM box diagram responsible for
the $K^0$--$\bar{K}^0$ mixing and take only two generations, i.e. only the up
and charm quarks run in the loop. In this case, the GIM mechanism yields a
suppression factor of ${\cal{O}}((m_c^2 - m_u^2)/M_W^2)$. If we replace the W
boson and the up quarks in the loop with their SUSY partners and we take,
for simplicity, all SUSY masses of the same order, we obtain a
super--GIM factor which looks like the GIM one with the masses of the
superparticles instead of those of the corresponding particles. The
problem is that the up  and charm squarks have masses 
which are much larger
than those of the corresponding quarks. Hence the super--GIM factor tends to
be of ${\cal{O}}(1)$ instead of being ${\cal{O}}(10^{-3})$ as it is in the SM
case. To
obtain this small number we would need a high degeneracy between the mass of
the charm and up squarks. It is difficult to think that such a degeneracy
may be accidental. After all, since we invoked SUSY for a naturalness
problem (the gauge hierarchy issue), we should avoid invoking
a fine--tuning to solve its problems! Then, one can turn to some symmetry
reason. For instance, just sticking to this simple example that we are
considering, one may think that the main bulk of the charm and up squark
masses is the same, i.e. the mechanism of SUSY breaking should have some
universality in providing the mass 
to these two squarks with the same
electric charge.  Flavor universality is by no means a prediction of
low--energy SUSY.
The absence of flavor universality of soft--breaking terms may result from
radiative effects at the GUT scale or from effective supergravities
derived 
from string theory. Indeed, from the point of view of these effective
supergravity theories, it may appear more natural
not to have such flavor universality. To obtain it one has to invoke
particular circumstances, like, for instance, strong dilaton over moduli
dominance in the breaking of supersymmetry, something which is certainly
not expected on general ground.

 Another possibility one may envisage is that the masses
of the squarks are quite high, say above few TeV's. Then, even if they are
not so degenerate in mass, the overall factor in front of the four--fermion
operator responsible for the kaon mixing becomes smaller and smaller (it
decreases quadratically with the mass of the squarks) and, consequently, one
can respect the observational result. We see from this simple example
that the issue of FCNC may be closely linked to the crucial problem of the
way we break SUSY.

We now turn to some general remarks about the worries and hopes that CP
violation arises in the SUSY context.

\vskip .8cm
\subsection{CP Violation in SUSY}

CP violation has major potentialities to exhibit manifestations of new physics
beyond the standard model.
Indeed, it is quite a general feature that new physics possesses
new CP violating phases in addition to the
Cabibbo--Kobayashi--Maskawa (CKM) phase $\left(\delta_{\rm CKM}\right)$
or, even in those cases where this does not occur, $\delta_{\rm CKM}$
shows up in interactions of the new particles, hence with potential departures
from the SM expectations. Moreover, although the SM is able to account for the
observed CP violation in the kaon system, we cannot say that we have tested so
far the SM predictions for CP violation. The detection of CP violation in $B$
physics will constitute a crucial test of the standard CKM picture within the
SM. Again, on general grounds, we expect new physics to provide departures from
the SM CKM scenario for CP violation in $B$ physics. A final remark on reasons
that make us optimistic in having new physics playing a major role in CP
violation concerns the matter--antimatter asymmetry in the universe. Starting
from a baryon--antibaryon symmetric universe, the SM is unable to account for
the observed baryon asymmetry. The presence of new CP--violating contributions
when one goes beyond the SM looks crucial to produce an efficient mechanism for
the generation of a satisfactory $\Delta$B asymmetry.

The above considerations apply well to the new physics represented by
low--energy supersymmetric extensions of the SM. Indeed, as we will see below,
supersymmetry introduces CP violating phases in addition to
$\delta_{\rm CKM}$ and, even if one envisages particular situations
where such extra--phases vanish, the phase $\delta_{\rm CKM}$ itself
leads to new CP--violating contributions in processes where SUSY particles are
exchanged. CP violation in $B$ decays has all potentialities to exhibit
departures from the SM CKM picture in low--energy SUSY extensions, although, as
we will discuss, the detectability of such deviations strongly depends on the
regions of the SUSY parameter space under consideration.

In any MSSM, at least two new ``genuine'' SUSY CP--violating phases are 
present. They
originate from the SUSY parameters $\mu$, $M$, $A$ and $B$. The first of these
parameters is the dimensionful coefficient of the $H_u H_d$ term of the
superpotential. The remaining three parameters are present in the sector that
softly breaks the N=1 global SUSY. $M$ denotes the common value of the gaugino
masses, $A$ is the trilinear scalar coupling, while $B$ denotes the bilinear
scalar coupling. In our notation, all these three parameters are
dimensionful. The simplest way to see which combinations of the phases of these
four parameters are physical \cite{Dugan} is to notice that for vanishing
values of $\mu$,  $M$, $A$ and $B$ the theory possesses two additional
symmetries \cite{Dimopoulos}. Indeed, letting $B$ and $\mu$ vanish, a $U(1)$
Peccei--Quinn symmetry originates, which in particular rotates $H_u$ and $H_d$.
If $M$, $A$ and $B$ are set to zero, the Lagrangian acquires a continuous
$U(1)$ $R$ symmetry. Then we can consider  $\mu$,  $M$, $A$ and $B$ as spurions
which break the $U(1)_{PQ}$ and $U(1)_R$ symmetries. In this way, the question
concerning the number and nature of the meaningful phases translates into the
problem of finding the independent combinations of the four parameters which
are invariant under $U(1)_{PQ}$ and $U(1)_R$ and determining their independent
phases. There are three such independent combinations, but only two of their
phases are independent. We use here the commonly adopted choice:
\begin{equation}
  \label{CMSSMphases}
  \varphi_A = {\rm arg}\left( A^* M\right), \qquad
  \varphi_B = {\rm arg}\left( B^* M\right).
\end{equation}
where also ${\rm arg}\left( B \mu\right) = 0$, i.e. 
$\varphi_\mu= - \varphi_B$.

The main constraints on $\varphi_A$ and $\varphi_B$ come from their 
contribution to
the electric dipole moments of the neutron and of the electron. For instance,
the effect of $\varphi_A$ and $\varphi_B$ on the electric and chromoelectric dipole
moments of the light quarks ($u$, $d$, $s$) lead to a contribution to
$d^e_N$ of 
order \cite{EDMN}
\begin{equation}
  \label{EDMNMSSM}
  d^e_N \sim 2 \left( \frac{100 {\rm GeV}}{\tilde{m}}\right)^2 \sin \varphi_{A,B}
  \times 10^{-23} {\rm e\, cm},
\end{equation}
where $\tilde{m}$ here denotes a common mass for squarks and gluinos. The
present experimental bound, $d^e_N < 1.1 \time 10^{-25}$ e cm, implies that
$\varphi_{A,B}$ should be $<10^{-2}$, unless one pushes SUSY masses up to 
${\cal{O}}$(1 TeV). A possible caveat to such an argument calling for a 
fine--tuning of
$\varphi_{A,B}$ is that uncertainties in the estimate of the hadronic matrix
elements could relax the severe bound in Eq.~(\ref{EDMNMSSM}) \cite{Ellis}.

In view of the previous considerations, most authors dealing with the MSSM
prefer to simply put $\varphi_A$ and $\varphi_B$ equal to zero. Actually, one may
argue in favor of this choice by considering the soft breaking sector of the
MSSM as resulting from SUSY breaking mechanisms which force $\varphi_A$ and
$\varphi_B$ to vanish. For instance, it is conceivable that both $A$ and $M$
originate from one same source of $U(1)_R$ breaking. Since $\varphi_A$ ``measures''
the relative phase of $A$ and $M$, in this case it would ``naturally''vanish. 
In some specific models, it has been shown \cite{Dine} that through an 
analogous mechanism also $\varphi_B$ may vanish.

If $\varphi_A=\varphi_B=0$, then the novelty of SUSY in CP violating 
contributions
merely arises from the presence of the CKM phase in loops where SUSY particles
run \cite{CPSUSY}. The crucial point is that the usual GIM suppression, which
plays a major role in evaluating $\varepsilon_K$ and 
$\varepsilon^\prime/\varepsilon$ in the SM, in the MSSM case (or more exactly 
in the CMSSM) is replaced by a super--GIM cancellation which has the same
``power'' of suppression as the original GIM (see previous section). Again,
also in
the CMSSM, as it is the case in the SM, the smallness of $\varepsilon_K$ and
$\varepsilon^\prime/\varepsilon$ is guaranteed not by the smallness of
$\delta_{\rm CKM}$, but
rather by the small CKM angles and/or small Yukawa couplings. By the same
token, we do not expect any significant departure of the CMSSM from the SM
predictions also concerning CP violation in $B$ physics. As a matter of fact,
given the large lower bounds on squark and gluino masses, one expects
relatively tiny contributions of the SUSY loops in $\varepsilon_K$ or
$\varepsilon^\prime/\varepsilon$ in comparison with the normal $W$ loops of 
the SM. Let us be more detailed on this point.

In the CMSSM, the gluino exchange contribution
to FCNC is subleading with respect to chargino ($\chi^\pm$) and charged
Higgs ($H^\pm$) exchanges. Hence, when dealing with CP violating FCNC
processes in the CMSSM with $\varphi_A=\varphi_B=0$, one can confine the analysis
 to $\chi^\pm$
and $H^\pm$ loops. If one takes all squarks to be degenerate in mass and
heavier than $\sim 200$ GeV, then $\chi^\pm-\tilde q$ loops are obviously
severely penalized with respect to the SM $W^+$--$q$ loops (remember that at the
vertices the same CKM angles occur in both cases).

The only chance for the CMSSM to produce some sizeable departure from the SM
situation in CP violation is in the particular region of the parameter space
where one has light $\tilde q$, $\chi^\pm$ and/or $H^\pm$. The best
candidate (indeed the only one unless 
$\tan \beta \sim m_t/m_b$) for a light squark is the stop. Hence one can
ask the following question: can the CMSSM present some novelties in CP--violating
phenomena when we consider $\chi^+$--$\tilde t$ loops with light $\tilde t$,
$\chi^+$ and/or $H^+$?

Several analyses in the literature tackle the above question or, to be more
precise, the more general problem of the effect of light $\tilde t$
and $\chi^+$ 
on FCNC processes \cite{refbrignole,mpr,branco}. A first important
observation concerns the
relative sign of the $W^+$--$t$ loop with respect to the  $\chi^+$--$\tilde t$ 
and $H^+$--$t$ contributions. As it is well known, the latter contribution 
always
interferes positively with the SM one. Interestingly enough, in the region of
the MSSM parameter space that we consider here, also the $\chi^+$--$\tilde t$
contribution interferes constructively with the SM contribution. The second
point regards the composition of the lightest chargino, i.e. whether the
gaugino or higgsino component prevails. This is crucial since the light stop is
predominantly $\tilde t_R$ and, hence, if the lightest chargino is mainly a
wino, it couples to $\tilde t_R$ mostly through the $LR$ mixing in the stop
sector. Consequently, a suppression in the contribution to box diagrams going
as $\sin^4 \theta_{LR}$ is present ($\theta_{LR}$ denotes the mixing angle
between
the lighter and heavier stops). On the other hand, if the lightest chargino is
predominantly a higgsino (i.e. $M_2 \gg \mu$ in the chargino mass matrix), then
the $\chi^+$--lighter $\tilde t$ contribution grows. In this case, 
contributions
$\propto \theta_{LR}$ become negligible and, moreover, it can be shown that
they are independent on the sign of $\mu$. A detailed study is provided in
reference \cite{mpr,branco}. For instance, for $M_2/\mu=10$, they find that 
the inclusion of
the SUSY contribution to the box diagrams doubles the usual SM contribution for
values of the lighter $\tilde t$ mass up to $100$--$120$ GeV, using $\tan \beta
=1.8$, $M_{H^+}=100$ TeV, $m_\chi=90$ GeV and the mass of the heavier $\tilde
t$ of 250 GeV. However, if $m_\chi$ is pushed up to 300 GeV, the  
$\chi^+$--$\tilde t$ loop yields a contribution which is roughly 3 times less 
than in the
case $m_\chi=90$ GeV, hence leading to negligible departures from the SM
expectation. In the cases where the SUSY contributions are sizeable, one
obtains relevant restrictions on the $\rho$ and $\eta$ parameters of the CKM
matrix by making a fit of the parameters $A$, $\rho$ and $\eta$ of the CKM
matrix and of the total loop contribution to the experimental values of
$\varepsilon_K$ and $\Delta M_{B_d}$. For instance, in the above--mentioned
case in which the SUSY loop contribution equals the SM $W^+$--$t$ loop, hence giving
a total loop contribution which is twice as large as in the pure SM case,
combining the $\varepsilon_K$ and $\Delta M_{B_d}$ constraints leads to a
region in the $\rho$--$\eta$ plane with $0.15<\rho<0.40$ and $0.18<\eta<0.32$,
excluding negative values of $\rho$.

In conclusion, the situation concerning CP violation in the MSSM case with
$\varphi_A=\varphi_B=0$ and exact universality in the soft--breaking sector can be
summarized in the following way: the MSSM does not lead to any significant
deviation from the SM expectation for CP--violating phenomena as $d_N^e$,
$\varepsilon_K$, $\varepsilon^\prime/\varepsilon$ and CP violation in $B$ physics; the only
exception to this statement concerns a small portion of the MSSM
parameter space 
where a very light $\tilde t$ ($m_{\tilde t} < 100$ GeV) and $\chi^+$
($m_\chi \sim 90$ GeV) are present. In this latter particular situation,
sizeable SUSY contributions to $\varepsilon_K$ are possible and, consequently,
major restrictions in the $\rho$--$\eta$ plane can be inferred. Obviously, CP
violation in $B$ physics becomes a crucial test for this MSSM case with very
light $\tilde t$ and $\chi^+$. Interestingly enough, such low values of SUSY
masses are at the border of the detectability region at LEP II.

In next Section, we will move to the case where, still keeping the
minimality of the model, we switch on the new CP violating phases.
Later on we will give up also the strict minimality related to the absence
of new flavor structure in the SUSY breaking sector and we will see that,
in those more general contexts, we can expect SUSY to significantly depart
from the SM predictions in  CP violating phenomena. 

\section{Flavor Blind SUSY Breaking and CP Violation}
\label{sec:flavor-blind}

We have seen in the previous section that in any MSSM there are additional 
phases which can cause deviations from the predictions of the SM in CP 
violation experiments. In fact, in the CMSSM, there are already two new phases 
present, Eq.(\ref{CMSSMphases}), and for most of the MSSM parameter 
space, 
the experimental bounds on the electric dipole moments (EDM) of the electron 
and neutron constrain these phases to be at most ${\cal{O}}(10^{-2})$.  
However, in the last few years, the possibility of having non--zero SUSY phases
has again attracted a great deal of attention. Several new mechanisms have 
been proposed to suppress supersymmetric contributions to EDMs below the 
experimental bounds while allowing SUSY phases ${\cal{O}}(1)$. 
Methods of suppressing the EDMs 
consist of cancellation of various SUSY contributions among themselves 
\cite{cancel}, non universality of the soft breaking parameters at the 
unification scale \cite{non-u} and approximately degenerate heavy sfermions 
for the first two generations \cite{heavy}. 
In the presence of one of these mechanisms, large supersymmetric phases are
naturally expected and EDMs should be generally close to the experimental 
bounds. \footnote{In a more general (and maybe more natural) MSSM
there are many other CP violating phases \cite{124} that contribute to CP 
violating observables.}

In this section we will study the effects of these phases in CP violation
observables as $\varepsilon_K$, $\varepsilon^\prime/\varepsilon$ and $B^0$ 
CP asymmetries. In particular we will show that the presence of large susy 
phases is not enough to produce sizeable supersymmetric contributions to 
these observables. In fact, {\it in the absence of the CKM phase, a general 
MSSM with all possible phases in the soft--breaking terms, but no new flavor 
structure beyond the usual Yukawa matrices, can never give a sizeable 
contribution to $\varepsilon_K$, $\varepsilon^\prime/\varepsilon$ or hadronic 
$B^0$ CP asymmetries}. However, we will see in the next section, that  
as soon as one introduces some new flavor structure in the soft Susy--breaking 
sector, even if the CP violating phases are flavor independent, it is indeed 
possible to get sizeable CP contribution for large Susy phases and 
$\delta_{CKM}=0$.
Then, we can rephrase our sentence above in a different way: {\it A new result 
in hadronic $B^0$ CP asymmetries in the framework of supersymmetry would be 
a direct proof of the existence of a completely new flavor structure in the 
soft--breaking terms}. This means that $B$--factories will probe the flavor 
structure of the supersymmetry soft--breaking terms even before the direct 
discovery of the supersymmetric partners \cite{flavor}. 

To prove this we will consider any MSSM, i.e. with the minimal supersymmetric 
particle content, with general {\bf complex} soft--breaking terms, but with a 
flavor structure strictly given by the two familiar Yukawa matrices or any 
matrix strictly proportional to them. In these conditions, the most general 
structure of the soft--breaking terms at the large scale, that we call 
$M_{GUT}$, is,
\begin{eqnarray}
\label{soft}
& (m_Q^2)_{i j} = m_Q^2\ \delta_{i j}\ \ \ 
(m_U^2)_{i j} = m_U^2\ \delta_{i j}\ \ \ 
(m_D^2)_{i j} = m_D^2\ \delta_{i j} &\nonumber \\
& (m_L^2)_{i j} = m_L^2\ \delta_{i j} \ \ \ 
(m_E^2)_{i j} = m_E^2\ \delta_{i j}\ \ \ \ m_{H_1}^2 \ \ \ \ \ m_{H_2}^2\ \ \
&\nonumber \\
& m_{\tilde{g}}\ e^{i \varphi_3}\ \ \ m_{\tilde{W}}\ e^{i \varphi_2}\ \ \ 
m_{\tilde{B}}\ e^{i \varphi_1} \ \ \ 
(A_U)_{i j}= A_U\ e^{i \varphi_{A_U}}\ (Y_U)_{i j}&\nonumber \\
& (A_D)_{i j}= A_D\ e^{i \varphi_{A_D}}\
(Y_D)_{i j}\ \ \ \ (A_E)_{i j}= A_E\ e^{i \varphi_{A_E}}\ (Y_E)_{i j}. & 
\end{eqnarray}
where all the allowed phases are explicitly written and one of them can be 
removed by an R--rotation. All other numbers or matrices in this equation 
are always real.
Notice that this structure covers, not only the CMSSM \cite{CPcons}, but also 
most of Type I string motivated models considered so far from phenomenology 
\cite{typeI,newcancel}, gauge mediated models \cite{GMSB1,GMSB2,GMSB3}, 
minimal effective supersymmetry models \cite{fully,CPbs}, etc.
 
Experiments of CP violation in the $K$ or $B$ systems only involve 
supersymmetric particles as virtual particles in the loops. This means that 
the phases in the soft--breaking terms can only appear in these experiments 
through the mass matrices of the SUSY particles. Then, the key point in our 
discussion will be the role played by the SUSY phases and the soft--breaking
terms flavor structure in the low--energy sparticle mass matrices.

It is important to notice that, even in a model with flavor--universal 
soft--breaking terms at some high energy scale, as this is the case, some 
off--diagonality 
in the squark mass matrices appears at the electroweak scale. Working on the 
the so--called Super CKM basis (SCKM), where squarks are rotated parallel to
the quarks so that Yukawa matrices are diagonalized, the squark mass matrix 
is not flavor diagonal at $M_W$. This is due to the fact that at $M_{GUT}$ 
there are always two 
non--trivial flavor structures, namely the two Yukawa matrices for the up and 
down quarks, not simultaneously diagonalizable. This implies that 
through RGE evolution some flavor mixing leaks into the sfermion mass matrices.
In a general Supersymmetric model, the presence of new flavor structures
in the soft breaking terms would generate large flavor mixing in the sfermion 
mass matrices. However, in the CMSSM, the two Yukawa matrices are the only 
source of flavor change. Always in the SCKM basis, any off--diagonal entry in 
the sfermion mass matrices at $M_W$ will be necessarily proportional to a 
product of Yukawa couplings.
Then, a typical estimate for the element $(i,j)$ in the $L$--$L$ down 
squark mass matrix at the electroweak scale would necessarily be (see 
\cite{CPcons} for details),
\begin{eqnarray}
\label{estimate1}
({m^2}_{LL}^{(D)})_{i j} \approx\ c\ m_Q^2\  Y^u_{i k} {Y_{j k}^u}^*,
\end{eqnarray}
with $c$ a proportionality factor between $0.1$ and 1.
This rough estimate provides the order of magnitude of the different entries
in the sfermion mass matrices. It is important to notice that if the phases 
of these elements were ${\cal O}(1)$, due to some of the phases in equation 
(\ref{soft}), we would be able to give sizeable contributions, or even 
saturate, the different CP observables \cite{gabbiani}. Then, it is clear that the 
relevant question for CP violation experiments is the presence of imaginary 
parts in these off--diagonal entries.

As explained in \cite{CPcons,BBM,RGE}, once we have solved the Yukawa RGEs, 
the RGE equations of all soft--breaking terms are a set of linear differential 
equations. Then, they can be solved as a linear function of the initial 
conditions. For instance the scalar masses are,
\begin{eqnarray}
\label{solution}
&m_{S}^{2}(M_{W})=\sum_i \eta^{(\phi_i)}_{S} m_{\phi_i}^{2} +
\sum_{i\neq j} \Big(\eta^{(g_i g_j )}_{S} e^{i(\varphi_{i}-
\varphi_j)} +\eta^{(g_i g_j)\,T}_{S} e^{- i(\varphi_{i}-\varphi_j)}\Big) m_{g_i}   m_{g_j}&
\nonumber\\
& +
\sum_i \eta^{(g_i)}_{S}\ m_{g_i}^{2}+\sum_{i j} \Big(\eta^{(g_i A_j )}_{S} 
e^{i(\varphi_{i}-\varphi_{A_j})} + \eta^{(g_i A_j)\,T}_{S} e^{- i(\varphi_{i}-\varphi_{A_j})}
\Big)  m_{g_i}   A_{j}&\nn \\ 
&+ \sum_i \eta^{(A_i)}_{S}  A_{i} ^{2} +
\sum_{i\neq j} \Big(\eta^{(A_i A_j )}_{S} e^{i(\varphi_{A_i}-\varphi_{A_j})}+
\eta^{(A_i A_j)\,T}_{S} e^{- i(\varphi_{A_i}-\varphi_{A_j})}\Big)  A_{i}   A_{j}& \nn\\
\end{eqnarray}
where $S=Q,U,D$, $\phi_i$ refers to any scalar, $g_i$ to the different 
gauginos and $A_i$ to any tri--linear coupling. In this equation, 
the different $\eta$ matrices are 
$3\times3$ matrices, {\bf strictly real} and all the allowed phases have been 
explicitly written. Regarding the imaginary parts, due to the hermiticity of
the sfermion mass matrices, any imaginary part will always be associated 
to the non--symmetric part of the $\eta^{(g_i g_j)}_{S}$, 
$\eta^{(A_i A_j )}_{S}$ or $\eta^{(g_i A_j)}_{S}$ matrices. 
To estimate the size of these anti--symmetric parts, we can go to the RGE 
equations for the scalar mass matrices, where we use the same conventions 
and notation as in \cite{CPcons,BBM}. 
Taking advantage of the linearity of these equations, we can directly write 
the evolution of the anti--symmetric parts, for instance $\hat{m}_{Q}^2 = 
m_{Q}^{2}-(m_{Q}^{2})^T$, as,
\begin{eqnarray}
\label{anti-sym}
\Frac{d \hat{m}_{Q}^{2}}{d t} =& - [ \Frac{1}{2}(\tilde{Y}_U \tilde{Y}_U^\dagger+\tilde{Y}_D 
\tilde{Y}_D^\dagger) \hat{m}_{Q}^{2} +\Frac{1}{2}\hat{m}_{Q}^{2}(\tilde{Y}_U \tilde{Y}_U^\dagger
+\tilde{Y}_D \tilde{Y}_D^\dagger) +\nonumber\\
& 2\ i\ \Im\{\tilde{A}_U \tilde{A}_U^\dagger + \tilde{A}_D 
\tilde{A}_D^\dagger\} + 
\tilde{Y}_U \hat{m}_{U}^{2} \tilde{Y}_U^\dagger + \tilde{Y}_D \hat{m}_{D}^{2} 
\tilde{Y}_D^\dagger ]
\end{eqnarray}
where, due to the reality of Yukawa matrices, we have used $Y^T = Y^\dagger$, 
and following \cite{BBM} a tilde over the couplings ($\tilde{Y}$, $\tilde{A}$, 
...) denotes a re--scaling by a factor $1/(4\pi)$.
The evolution of the $R$--$R$ squark mass matrices, $m_U^2$ and $m_D^2$, 
is completely analogous. With the initial conditions in equation (\ref{soft}),
$\hat{m}_{Q}^2$, $\hat{m}_{U}^2$ and $\hat{m}_{D}^2$ at $M_{GUT}$ are 
identically zero. Then, we can safely 
neglect the last two terms in equation (\ref{anti-sym}) because they will 
only be a second order effect.
This means that the only source for $\hat{m}_{Q}^{2}$ in equation 
(\ref{anti-sym}) is necessarily $\Im\{A_U A_U^\dagger + A_D A_D^\dagger\}$
(also for $\hat{m}_{U,D}^2$) .

The next step is then to analyze the RGE for the tri--linear couplings,
\begin{eqnarray}
\label{Aurge}
&\Frac{d \tilde{A}_U}{d t} = \Frac{1}{2} \Big(\Frac{16}{3}\tilde{\alpha}_3 + 
3 \tilde{\alpha}_2 + \Frac{1}{9}\tilde{\alpha}_1 \Big) \tilde{A}_U -
\Big(\Frac{16}{3}\tilde{\alpha}_3 M_3 + 3 \tilde{\alpha}_2 M_2 + 
\Frac{1}{9}\tilde{\alpha}_1 M_1\Big) \tilde{Y}_U - &\nonumber
\\
&\Big( 2\tilde{A}_U \tilde{Y}_U^\dagger\tilde{Y}_U + 
3 Tr(\tilde{A}_U \tilde{Y}_U^\dagger)\tilde{Y}_U 
+ \Frac{5}{2} \tilde{Y}_U \tilde{Y}_U^\dagger \tilde{A}_U + 
\Frac{3}{2} Tr(\tilde{Y}_U \tilde{Y}_U^\dagger) \tilde{A}_U +& \nonumber \\ 
&\tilde{A}_D \tilde{Y}_D^\dagger\tilde{Y}_U + 
\Frac{1}{2}\tilde{Y}_D \tilde{Y}_D^\dagger \tilde{A}_U \Big)&
\end{eqnarray}
with an equivalent equation for $A_D$. With the general 
initial conditions in equation (\ref{soft}), $A_U$ is complex at any scale. 
However, we are interested in the imaginary parts of $A_U A_U^\dagger$. 
At $M_{GUT}$ this combination is exactly real, but, due to different 
renormalization of different elements of the matrix, this is not
true anymore at a different scale. 
Nevertheless, a careful analysis of equation (\ref{Aurge}) is enough to 
convince ourselves that these imaginary parts are extremely small. 
Let us, for a moment, neglect the terms involving 
$\tilde{A}_D \tilde{Y}_D^\dagger$ or $\tilde{Y}_D \tilde{Y}_D^\dagger$ 
from the above equation or, strictly speaking, from the complete set of MSSM
RGE. Then, the only flavor structure appearing in 
equation (\ref{Aurge}) at $M_{GUT}$ is $Y_U$. We can always go to the basis 
where $Y_U$ is diagonal and then we will have $A_U$ exactly diagonal at 
any scale. In particular this means that  $\Im\{A_U A_U^\dagger\}$ would 
always exactly vanish. A completely parallel reasoning can be applied to 
$A_D$ and $\Im\{A_D A_D^\dagger\}$. Hence, simply taking into account the 
flavor structure, our conclusion is that, necessarily, any non--vanishing 
element of $\Im[A_U A_U^\dagger + A_D A_D^\dagger]$ and hence of 
$\hat{m}_{Q}^{2}$ must be proportional to $(\tilde{Y}_D \tilde{Y}_D^\dagger 
\tilde{Y}_U \tilde{Y}_U^\dagger - H.C.)$.     
So, we can expect them to be,
\begin{eqnarray}
\label{im-estimate}
& (\hat{m}_{Q}^{2})_{i<j} \approx K \left(Y_D Y_{D}^{\dagger} 
Y_U Y_{U}^{\dagger} - H.C.\right)_{i<j}& \nonumber \\
&(\hat{m}_{Q}^{2})_{1 2} \approx K \cos^{-2}\beta~ (h_{s} h_t \lambda^{5})& 
\nonumber \\
&(\hat{m}_{Q}^{2})_{1 3} \approx K \cos^{-2}\beta~ (h_{b} h_t \lambda^{3})&
\nonumber\\ 
&(\hat{m}_{Q}^{2})_{2 3} \approx K \cos^{-2}\beta~ (h_{b} h_t \lambda^{2}),&
\end{eqnarray}
where $h_{i}=m_{i}^{2}/v^2$, with $v=\sqrt{v_1^2+v_2^2}$ the vacuum 
expectation value of the Higgs, $\lambda=\sin \theta_c$ and $K$ is a 
proportionality constant that includes the effects of the running from 
$M_{GUT}$ to $M_W$. To estimate this constant, we have to keep in mind that 
the imaginary parts of $A_U A_U^\dagger$ are generated through the RGE running 
and then, these imaginary parts generate $\hat{m}_{Q}^{2}$ as a second order 
effect. This means that roughly $K \simeq {\cal O}(10^{-2})$ times a 
combination of initial conditions as in equation (\ref{solution}). So, 
we estimate these matrix elements to be $ (\cos^{-2} \beta \{ 10^{-12}, 
6 \times 10^{-8}, 3 \times 10^{-7}\})$ times initial conditions.
This was exactly the result we found for the $A$--$g$ terms in \cite{CPcons} 
in the framework of the CMSSM. 
In fact, now it is clear that this is the same for all the terms in 
equation (\ref{solution}), $g_i$--$A_j$,  $g_i$--$g_j$ and $A_i$--$A_j$, 
irrespectively of the presence of an arbitrary number of new phases.
This discussion can be directly applied for the $R$--$R$ matrices. 

Hence, so far, we have shown that the $L$--$L$ or $R$--$R$ squark mass 
matrices are still essentially real.
The only complex matrices, then, will still be the $L$--$R$ matrices that 
include, from the very beginning, the phases $\varphi_{A_i}$ and 
$\varphi_\mu$. Once more, the size of these entries is determined by the 
Yukawa elements with these two phases providing the complex structure.
In fact, we can follow the same reasoning used after Eq.(\ref{Aurge}). 
In the absence of the terms involving $\tilde{A}_D \tilde{Y}_D^\dagger$ or 
$\tilde{Y}_D \tilde{Y}_D^\dagger$, the $\tilde{A}_U$ matrix would be exactly 
diagonalized when we diagonalize the Yukawa matrices. So, any off--diagonal 
element in the SCKM basis will be proportional to three Yukawas, 
$(\tilde{Y}_D \tilde{Y}_D^\dagger \tilde{Y}_U)$ and hence sufficiently small.  
Notice that this situation is not new for these more general MSSM models and 
it was already present even in the CMSSM. We can conclude, then, that the 
structure of the sfermion mass matrices at $M_W$ is not modified from the 
familiar structure already present in the CMSSM, irrespective of the presence 
of an arbitrary number of new SUSY phases.

\subsection{Indirect CP violation}
\label{sec:indirect}

Next, we will consider indirect CP violation both in the $K$ and 
$B$ systems. In the SM, neutral meson mixing arises at one loop through the 
well--known $W$--box. However, in the MSSM, there are new contributions to 
$\Delta F=2$ processes coming from boxes mediated by supersymmetric particles. 
These are: charged Higgs boxes ($H^{\pm}$), chargino boxes ($\chi^{\pm}$) and 
gluino--neutralino boxes ($\tilde{g}$, $\chi^{0}$). ${\cal M}$--$\bar{\cal M}$ 
mixing is correctly described by the $\Delta F=2$ effective Hamiltonian, 
${\cal{H}}_{eff}^{\Delta F=2}$, which can be decomposed as,
\begin{eqnarray}
\label{DF=2}
{\cal{H}}_{eff}^{\Delta F=2}= - \Frac{G_{F}^{2} M_{W}^{2}}{(2 \pi)^{2}} 
(K_{td}^{*} K_{tq})^{2} ( C_{1}(\mu) Q_{1}(\mu) + C_{2}(\mu)  Q_{2}(\mu)  
+ C_3(\mu) Q_3(\mu)).\nonumber \\
\end{eqnarray}
With the relevant four--fermion operators given by 
\begin{eqnarray}
\label{ops}
Q_{1}=\bar{d}^{\alpha}_{L}\gamma^{\mu}q^{\alpha}_{L}\cdot 
\bar{d}^{\beta}_{L}\gamma_{\mu}q^{\beta}_{L},\ \  
Q_{2}=\bar{d}^{\alpha}_{L}q^{\alpha}_{R}\cdot \bar{d}^{\beta}_{L}
q^{\beta}_{R},\ \ 
Q_{3}=\bar{d}^{\alpha}_{L}q^{\beta}_{R}\cdot \bar{d}^{\beta}_{L}
q^{\alpha}_{R},
\end{eqnarray}
where $q=s , b$ for the $K$ and $B$--systems respectively, $\alpha, 
\beta$ are color indices and $K_{i j}$ is the CKM mixing matrix. 
In the CMSSM, these are the only three operators
present in the limit of vanishing $m_d$. The Wilson coefficients, $C_1(\mu)$,
$C_2(\mu)$ and $C_3(\mu)$, receive contributions from the different 
supersymmetric boxes,
\begin{eqnarray}
\label{wilson}
{C_{1}}(M_W)&=&{C_{1}^{W}}(M_W)+{C_{1}^{H}}(M_W)+ 
C_{1}^{\tilde{g},\chi^0}(M_W)+ {C_{1}^{\chi}}(M_W)\,\nonumber \\
{C_{2}}(M_W)&=&{C_{2}^{H}}(M_W) + C_{2}^{\tilde{g}}(M_W)\nonumber \\
{C_{3}}(M_W)&=& C_{3}^{\tilde{g},\chi^0}(M_W)+ {C_{3}^{\chi}}(M_W)
\end{eqnarray} 

Both, the usual SM $W$--box and the charged Higgs box contribute to these 
operators. However, with $\delta_{CKM}=0$, these contributions do not contain 
any complex phase and hence cannot generate an imaginary part for these
Wilson coefficients.

Then, Gluino and neutralino contributions are specifically supersymmetric. 
They involve the superpartners of quarks and gauge bosons. Here, the source of 
flavor mixing is not directly the usual CKM matrix. It is the presence of
off--diagonal elements in the sfermion mass matrices, as discussed in section
\ref{sec:flavor-blind}. To analyze these contributions, it is convenient to 
use the so--called Mass Insertion (MI) approximation \cite{MI,gabbiani}.  
To define the MI we go to the SCKM basis. In this basis, off--diagonal
flavor--changing effects can be estimated by insertion of 
flavor--off--diagonal components of the mass--squared matrices.  
By normalizing by an average squark mass--squared $m^{2}_{\tilde{q}}$, 
we define, 
\begin{eqnarray}
\label{MIns}
(\delta^{d}_{LL})_{ij}=\Frac{({m^{2}}^{(d)}_{LL})_{ij}}{m^{2}_{\tilde{q}}}\ \
(\delta^{d}_{RR})_{ij}=\Frac{({m^{2}}^{(d)}_{RR})_{ij}}{m^{2}_{\tilde{q}}}\ \
(\delta^{d}_{LR})_{ij}=\Frac{({m^{2}}^{(d)}_{LR})_{ij}}{m^{2}_{\tilde{q}}}
\end{eqnarray}
with ${m^{2}}^{(d)}_{AB}$ the squark mass matrices in the SCKM basis.

From the point of view of CP violation, we will always need a complex Wilson 
coefficient. In the SCKM basis all gluino vertices are flavor diagonal and 
real. Then, a complex MI in one of the sfermion lines is always required.
Only $L$--$L$ mass insertions enter at first order in the Wilson coefficient 
${C_{1}}^{\tilde{g},\chi^0}(M_W)$. From equation (\ref{im-estimate}), the 
imaginary parts of these MI are at most ${\cal O}(10^{-6})$ for the $b$--$s$ 
transitions and smaller otherwise \cite{CPcons}. 
Comparing these values with the phenomenological bounds required to saturate 
the measured values of these processes \cite{gabbiani} we can easily see that 
we are always several orders of magnitude below.

In the case of the Wilson coefficients ${C_{2}}^{\tilde{g}}(M_W)$ and 
${C_{3}}^{\tilde{g}}(M_W)$, the involved MI are $L$--$R$. However, as 
explained in
section \ref{sec:flavor-blind}, these MI are always 
suppressed by light masses of right handed squark plus two additional up 
Yukawas. Moreover, in the case of $b$--$s$ transitions they are directly 
constrained by the $b \rightarrow s \gamma$ decay. Hence,
gluino boxes, in the absence of new flavor structures, can never give 
sizeable contributions to indirect CP violation processes \cite{CPcons}.

The chargino contributions to these Wilson coefficients were 
discussed in great detail in the CMSSM framework in reference \cite{CPcons}. 
In this more general MSSM, we find very 
similar results due to the absence of new flavor structure.

Basically, in the chargino boxes, flavor mixing comes explicitly from the 
CKM mixing matrix, although off--diagonality in the sfermion mass matrix 
introduces a small additional source of flavor mixing.  
\begin{eqnarray}
\label{chWC}
&C_1^\chi (M_W) = \sum_{i,j=1}^{2} \sum_{k, l=1}^{6} 
\sum_{\alpha \gamma \alpha^\prime \gamma^\prime} \Frac{K_{\alpha^\prime d}^{*} 
K_{\alpha q}K_{\gamma^\prime d}^{*} K_{\gamma q}}{(K_{td}^{*} K_{tq})^2}&
\nonumber \\
&[G^{(\alpha,k)i} G^{(\alpha^\prime,k)j*} G^{(\gamma^\prime,l)i*} 
G^{(\gamma,l)j}\  Y_1(z_{k}, z_{l}, s_i, s_j)]& 
\end{eqnarray}
where $K_{\alpha q} G^{(\alpha,k)i}$ represent the coupling of chargino and 
squark $k$ to left--handed down quark $q$, $z_k = M_{\tilde{u}_k}^2/M_W^2$ 
and $s_i =M_{\tilde{\chi}_i}^2/M_W^2$. 
The explicit expressions for the loop functions can be found in reference 
\cite{CPcons}. These couplings, in terms of the standard mixing matrices
\cite{haber,BBM},
\begin{eqnarray}
G^{(\alpha, k) i}&=& \left( \Gamma_{U L}^{k \alpha} V_{i 1}^{*} - 
\Frac{m_{\alpha}}{\sqrt{2} M_W \sin \beta} \Gamma_{U R}^{k \alpha} 
V_{i 2}^{*}\right).
\end{eqnarray}  
$G^{(\alpha, k) i}$ are in general complex, as both $\varphi_\mu$ and 
$\varphi_{A_i}$ are present in the different mixing matrices. 

\begin{figure}
\begin{center}
\epsfxsize = 13cm
\epsffile{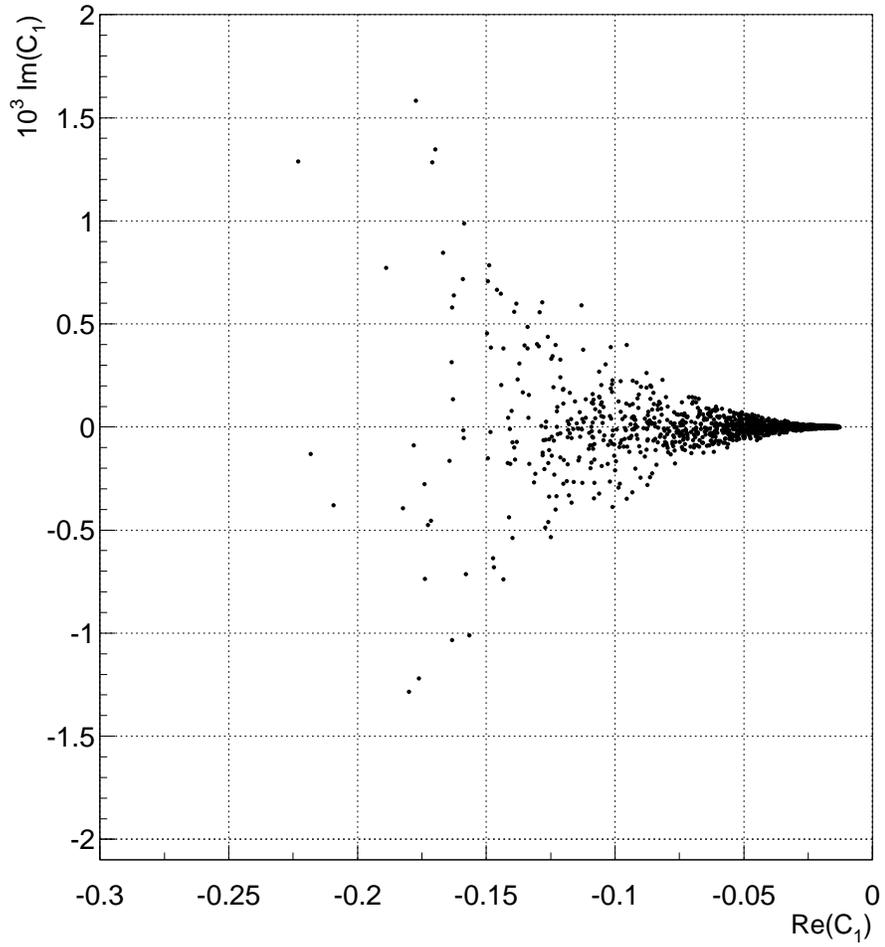}
\leavevmode
\end{center}
\caption{Imaginary and Real parts of the Wilson coefficient $C_1^\chi$ in
B mixing.}
\label{imB40}
\end{figure}

The main part of $C_1^\chi$ in equation (\ref{chWC}) will be given by pure CKM 
flavor mixing, neglecting the additional flavor mixing in the squark
mass matrix \cite{cho,branco}. This means, $\alpha= \alpha^\prime$ and
$\gamma=\gamma^\prime$. In these conditions, using the symmetry of the loop 
function $Y_1(a, b, c, d)$ under the exchange of any two indices it is easy to
prove that $C_1^\chi$ would be exactly real \cite{fully}. 
This is not exactly true either in the CMSSM or in our more general MSSM, 
where there is additional flavor change in the sfermion mass matrices. 
Here, some imaginary parts appear in the $C_1^\chi$ in equation (\ref{chWC}). 
In figure \ref{imB40} we show in a scatter plot the size of imaginary 
and real parts of $C_1^\chi$ in the B system for a fixed value of 
$\tan \beta=40$. We see that this Wilson coefficient is always real up to 
a part in $10^3$. In any case, this is out of reach for the foreseen 
B--factories. For the K system, imaginary parts are still smaller due to 
smaller mixing angles with the stop.

Finally, chargino boxes contribute also to the quirality changing Wilson 
coefficient $C_3^\chi(M_W)$,
\begin{eqnarray}
\label{chWCR}
&C_3^\chi (M_W) = \sum_{i,j=1}^{2} \sum_{k, l=1}^{6} \sum_{\alpha \gamma
\alpha^\prime \gamma^\prime}
\Frac{K_{\alpha^\prime d}^{*} K_{\alpha q}K_{\gamma^\prime d}^{*} 
K_{\gamma q}}{(K_{td}^{*} K_{tq})^2}\  \Frac{m_q^2}{2 M_W^2 \cos^2 \beta}&
\nonumber \\ 
&H^{(\alpha,k)i} G^{(\alpha^\prime,k)j*} G^{(\gamma^\prime,l)i*}  
H^{(\gamma,l)j} Y_2(z_k, z_l, s_i, s_j)& 
\end{eqnarray} 
where $m_q/(\sqrt{2} M_W \cos \beta) \cdot K_{\alpha q} \cdot H^{(\alpha,k)i}$ 
is the coupling of chargino and squark to the right--handed down quark $q$
\cite{haber,BBM},
\begin{equation}
\label{couplingR}
H^{(\alpha,k)i} = - U_{i 2} \Gamma_{U L}^{k \alpha}.
\end{equation} 
Unlike the $C_1^\chi$ Wilson coefficient, due to the differences between $H$ 
and $G$ couplings, $C_3^\chi$ is complex even in the 
absence of intergenerational mixing in the sfermion mass matrices \cite{fully}.
Then, the presence of these small flavor violating entries in the up--squark 
mass matrix hardly modifies the results obtained in their absence 
\cite{cho,branco,CPcons}.
In fact, in spite the presence of the Yukawa coupling squared, 
$m_q^2/(2 M^2_W \cos^2 \beta)$, this contribution could be relevant
in the large $\tan \beta$ regime. For instance, in $B^0$--$\bar{B}^0$ mixing
we have $m_b^2/(2 M^2_W \cos^2 \beta)$ that for $\tan \beta \gsim 25$ is
larger than 1 and so, it is not suppressed at all when compared with the 
$C_1^\chi$ Wilson Coefficient. This means that this contribution can be very 
important in the large $\tan \beta$ regime \cite{fully} and could have 
observable effects in CP violation experiments in the new B--factories.   
However, we will show next that when we include the constraints coming 
from $b \rightarrow s \gamma$ these chargino contributions are also reduced 
to an unobservable level.

The chargino contributes to the $b \rightarrow s \gamma$ decay through the
Wilson coefficients ${\cal{C}}_{7}$ and ${\cal{C}}_{8}$, corresponding to the
photon and gluon dipole penguins respectively \cite{BBM,bsg,CPcons}.
In the large $\tan \beta$ regime, we can 
approximate these Wilson coefficients as \cite{CPcons},
\begin{eqnarray}
&{\cal{C}}_{7}^{\chi^{\pm}}(M_W)=\sum_{k=1}^{6}\sum_{i=1}^{2}
\sum_{\alpha, \beta =u,c,t}\ \Frac{ K_{\alpha b} K_{\beta s}^{*}}
{K_{t b} K_{t s}^{*}}\Frac{m_{b}}{\sqrt{2} M_W \cos \beta}\ & \nonumber\\
& H^{(\alpha, k) i}
{G^{*}}^{(\beta, k) i}\ \Frac{M_{\chi^{i}}}{m_{b}}\ F_{R}^{7}(z_{k},s_{i})&
\nonumber \\
&{\cal{C}}_{8}^{\chi^{\pm}}(M_W)=\sum_{k=1}^{6}\sum_{i=1}^{2}
\sum_{\alpha, \beta =u,c,t}\ \Frac{ K_{\alpha b} K_{\beta s}^{*}}
{K_{t b} K_{t s}^{*}}\Frac{m_{b}}{\sqrt{2} M_W \cos \beta}\ & \nonumber \\
& H^{(\alpha, k) i}
{G^{*}}^{(\beta, k) i}\ \Frac{M_{\chi^{i}}}{m_{b}}\ F_{R}^{8}(z_{k},s_{i})
\label{charex}
\end{eqnarray}

Now, if we compare the chargino contributions to these Wilson coefficients
and to the coefficient $C_3$, equations (\ref{chWCR}) and (\ref{charex}), 
we can see that they are deeply related. In fact, in the approximation where
the two different loop functions involved are of the same order, we have,
\begin{eqnarray}
\label{approx}
C_3(M_W) \approx ({\cal C}_7(M_W))^2 \Frac{m_q^2}{M_W^2}
\end{eqnarray}

\begin{figure}
\vspace{9pt}
\begin{center}
\epsfxsize = 13cm
\epsffile{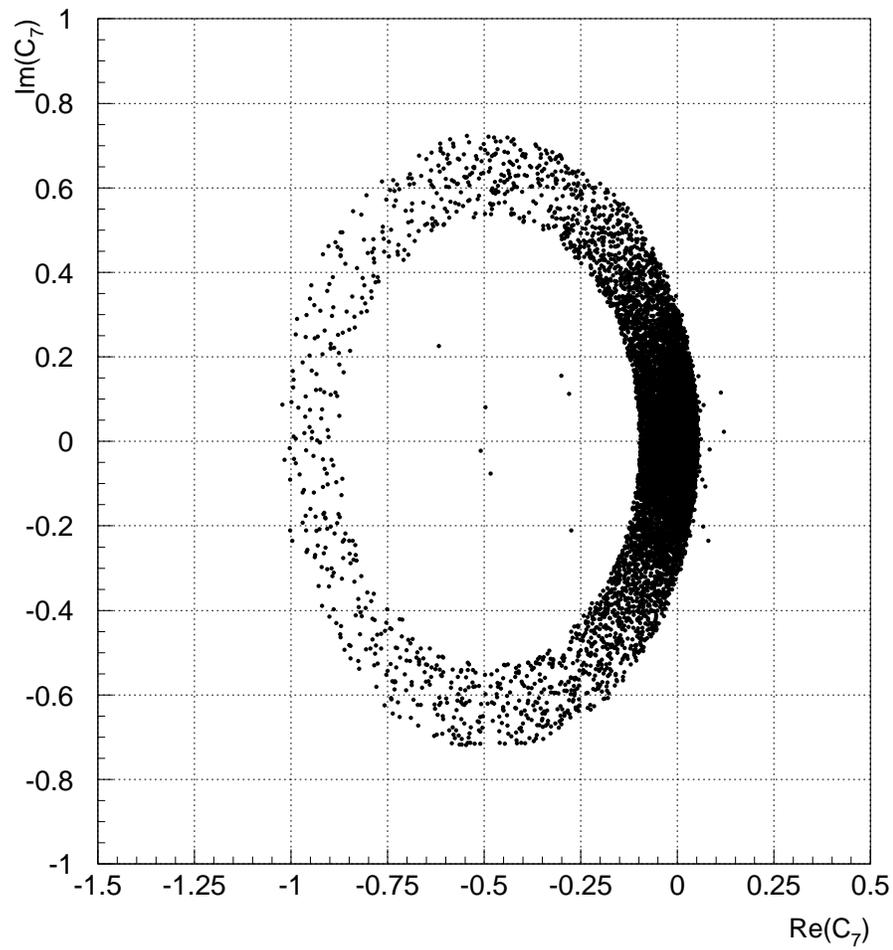}
\leavevmode
\end{center}
\caption{Experimental constraints on the Wilson Coefficient ${\cal C}_7$ }
\label{figc7}
\end{figure}

\begin{figure}
\begin{center}
\epsfxsize = 13cm
\epsffile{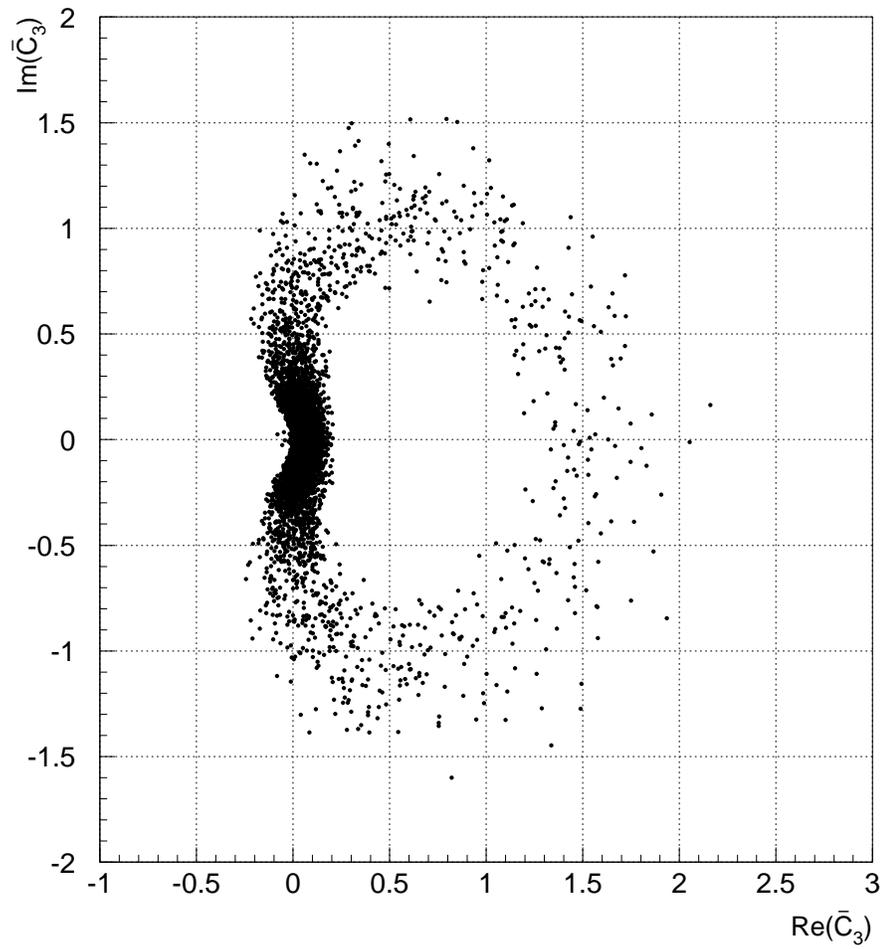}
\leavevmode
\end{center}
\caption{Allowed values for the re--scaled WC $\bar{C}_3$}
\label{figc3}
\end{figure}

In figure \ref{figc7}, we show a scatter plot of the allowed values of 
$Re({\cal C}_{7})$ versus $Im({\cal C}_{7})$ in the CMSSM for a fixed value of 
$\tan \beta = 40$ \cite{CPcons} with the constraints from the decay 
$B\rightarrow X_{s} \gamma $ taken from the reference \cite{kagan-neubert}. 
Notice that a relatively large value of $\tan \beta$, for example, 
$\tan \beta \gsim 10$, is needed to compensate the $W$ and charged Higgs 
contributions and cover the whole allowed area with positive and negative 
values. However, the shape of the plot is clearly independent of $\tan \beta$, 
only the number of allowed points and its location in the allowed area depend 
on the value considered. Then, figure \ref{figc3} shows the allowed values 
for a re--scaled Wilson coefficient $\bar{C}_3(M_W)= M^2_W/m_q^2 C_3(M_W)$ 
corresponding to the same allowed points of the SUSY parameter space in 
figure \ref{figc7}.
As we anticipated previously, the allowed values for $\bar{C}_3$ are close 
to the square of the values of ${\cal C}_7$ in figure \ref{figc7} slightly 
scaled by different values of the loop functions. 

We can immediately translate this result to a constraint on the size of the 
chargino contributions to $\varepsilon_{\cal M}$. 
\begin{eqnarray}
\label{epscoef}
&\varepsilon_{\cal M}\ =\ \Frac{G_F^2 M_W^2}{4 \pi^2 \sqrt{2}\ \Delta 
M_{\cal M}}\ \Frac{(K_{td} K_{tq})^2}{24}
F_{\cal M}^2\  M_{\cal M}\ &\nonumber \\& \Frac{M^2_{\cal M}}{m_q^2(\mu)+ 
m_d^2(\mu)}\ \eta_3(\mu)\  B_3(\mu)\  Im[C_3]&
\end{eqnarray}
In this expression $M_{{\cal{M}}}$, $\Delta M_{\cal M}$ and $F_{{\cal{M}}}$ 
denote the mass, mass difference and decay constant of the neutral meson 
${\cal{M}}^{0}$. The coefficient $\eta_3(\mu)= 2.93$ \cite{ciuchini} includes 
the RGE effects from $M_W$ to the meson mass scale, $\mu$, and $B_3(\mu)$ is 
the B--parameter associated with the matrix element of the $Q_3$ operator
\cite{ciuchini}.
  
For the $K$ system, using the experimentally measured value 
of $\Delta M_{K}$ we obtain,
\begin{eqnarray}
\label{epsK}
& \varepsilon_K^\chi\ =\ 1.7 \times 10^{-2}\ \Frac{m_s^2}{M_W^2}\ 
Im[\bar{C}_3] \approx 0.4  \times  10^{-7}\  Im[\bar{C}_3]&
\end{eqnarray}
Given the allowed values of $\bar{C}_3$ in figure \ref{figc3}, this means that
in the MSSM, even with large SUSY phases, chargino cannot produce a sizeable 
contribution to $\varepsilon_K$. 

The case of $B^0$--$\bar{B}^0$ mixing has a particular interest due to the 
arrival of new data from the B--factories. In fact, in the large $\tan \beta$ 
regime chargino contributions to indirect CP violation can be very important.
However, for any value of $\tan \beta$, we must satisfy the bounds from the
$b \rightarrow s \gamma$ decay. Then, if we apply these constraints to the 
$B^0$--$\bar{B}^0$ mixing,
\begin{eqnarray}
\label{epsB}
& \varepsilon_B^\chi\  =\  0.17\  \Frac{m_b^2}{M_W^2}\  Im[\bar{C}_3]
\approx 0.5 \times 10^{-3}\  Im[\bar{C}_3]&
\end{eqnarray}
where once again, with the allowed values of figure \ref{figc3} we get a very 
small contribution to CP violation in the mixing.
We must take into account that the mixing--induced CP phase, $\theta_M$, 
measurable in $B^0$ CP asymmetries, is related to $\varepsilon_B$ by 
$\theta_M=\arcsin\{2 \sqrt{2} \cdot \varepsilon_B \}$. The expected 
sensitivities on the CP phases at the B factories are around $\pm 0.1$ 
radians, so this supersymmetric chargino contribution will be absolutely 
out of reach.

\subsection{Direct CP violation}
\label{sec:direct}

To complete our analysis, we consider now direct CP violation. In this
case, the different decay processes are described by a $\Delta F=1$ 
effective Hamiltonian. A complete operator basis for these
transitions in a general MSSM involves 14 different operators \cite{gabbiani}.
The main difference with the case of indirect CP violation is that these
operators receive contributions both from box and penguin diagrams.
Nevertheless, the discussion of the presence of imaginary parts is completely
analogous to the case of indirect CP violation.

Once more, in the gluino case, $L$--$L$ transitions are real to a 
very good approximation, and several orders of magnitude below
the phenomenological bounds \cite{gabbiani}. On the other hand, 
$L$--$R$ transitions
are suppressed by two up Yukawas and a down quark mass and 
$b \rightarrow s \gamma$ decay. This is always true for the squark mass 
matrices obtained in section \ref{sec:flavor-blind}, and valid both for boxes 
and penguins.
 
Finally, we are left with chargino contributions. The analysis of chargino 
boxes is exactly the same as in the previous section. In fact, even the 
Wilson coefficients are identical except some CKM elements that can always be
factored out.
Then, for the penguins, $L$--$L$ transitions are exactly real if we neglect
inter--generational mixing in the squark mass matrices. Taking into account
this small mixing we find, for the very same reasons as in the indirect CP 
violation case, that imaginary parts are far too small.
The relation of the $b \rightarrow s \gamma$ decay with the $L$--$R$ chargino 
penguins is in this case even more transparent than for the boxes.
So, our conclusion is again that no new supersymmetric CP violation effects 
are possible in $\varepsilon^\prime/\varepsilon$ or hadronic $B$ CP 
asymmetries. 

However, there is still one possibility to observe the effects of the new 
supersymmetric phases even in the absence of new flavor structure. 
We have seen that the reason 
for the smallness of the contributions of chargino $L$--$R$ transitions is
the experimental bound from the $B\rightarrow X_{s} \gamma$ branching ratio.
This bound makes the chirality changing transitions, although complex, too 
small to compete with $L$--$L$ transitions. Hence, in these conditions,
just the processes where only
chirality changing operators contribute (EDMs or $b\rightarrow s \gamma$), or
observables where chirality flip operators are relevant ($b\rightarrow s l^+
l^-$) can show the effects of new supersymmetric phases \cite{CPbs,CPcons}.
 
\section{CP Violation in the presence of new Flavor Structures}
\label{sec:newflavor}
In section \ref{sec:flavor-blind}, we have shown that CP violation effects are 
always small in models with flavor blind soft--breaking terms. 
However, as soon as one introduces some new flavor structure in the soft 
breaking sector, it is indeed possible to get sizeable CP contribution 
for large Susy phases and $\delta_{CKM}=0$ \cite{non-u,brhlik2,newflavor}.
To show this, we will mainly concentrate in new supersymmetric 
contributions to $\varepsilon^\prime/\varepsilon$.
 
In the CMSSM, the SUSY contribution to $\varepsilon^\prime/\varepsilon$ is
small \cite{giudice,flavor}. However in a MSSM with a more general framework of
flavor structure it is relatively easy to obtain larger SUSY effects to
$\varepsilon^\prime/\varepsilon$. In ref. \cite{murayama} it was shown
that such large SUSY contributions arise once one assumes that:  i)
hierarchical quark Yukawa matrices are protected by flavor symmetry, ii) a
generic dependence of Yukawa matrices on Polonyi/moduli fields is present
(as expected in many supergravity/superstring theories), iii) the Cabibbo
rotation originates from the down--sector and iv) the phases are of order
unity. In fact, in \cite{murayama}, it was illustrated how the observed
$\varepsilon^\prime/\varepsilon$ could be mostly or entirely due to the
SUSY contribution. 

The universality of the breaking is a strong assumption and is known not
to be true in many supergravity and string inspired models \cite{BIM}.
In these models, we expect at least some non--universality in the squark
mass matrices or tri--linear terms at the supersymmetry breaking scale.
Hence, sizeable flavor--off-diagonal entries will appear in the squark
mass matrices.
In this regard, gluino contributions to $\varepsilon^\prime/\varepsilon$
are especially sensitive to $(\delta^{d}_{12})_{LR}$; even 
$|{\rm Im}(\delta^{d}_{12})_{LR}^{2}| \sim 10^{-5}$ gives a significant 
contribution to $\varepsilon^\prime/\varepsilon$ while keeping the
contributions from this MI to $\Delta m_{K}$ and $\varepsilon_K$ well bellow 
the phenomenological bounds. The situation is the opposite for $L$--$L$
and $R$--$R$ mass insertions; the stringent bounds on 
$(\delta^{d}_{12})_{LL}$ and $(\delta^{d}_{12})_{RR}$ 
from $\Delta m_{K}$ and $\varepsilon_K$ prevent them to contribute 
significantly to $\varepsilon^\prime/\varepsilon$. 

The LR squark mass matrix has the same flavor structure as the fermion 
Yukawa matrix and both, in fact, originate from the superpotential couplings. 
It may be appealing to invoke the presence of an underlying flavor symmetry 
restricting 
the form of the Yukawa matrices to explain their hierarchical forms. Then, 
the LR mass matrix is expected to have a very similar form as the Yukawa
matrix. Indeed, we expect the components of the LR mass matrix to be
roughly the SUSY breaking scale (e.g., the gravitino mass) times the
corresponding component of the quark mass matrix. However, there is no
reason for them to be simultaneously diagonalizable based on this general
argument.  
To make an order of magnitude estimate, we take the down quark mass matrix 
for the first and second generations to be (following our assumption iii)),
\begin{equation}
        Y^{d} v_1 \simeq \left( \begin{array}{cc}
                m_{d} & m_{s} V_{us} \\
                 & m_{s}
        \end{array} \right),
\end{equation}
where the (2,1) element is unknown due to our lack of knowledge on 
the mixings among right--handed quarks (if we neglect small terms 
$m_d V_{cd}$).  Based on the general 
considerations on the LR mass matrix above, we expect
\begin{equation}
        {m^{2}}^{(d)}_{LR} \simeq m_{3/2}  \left( \begin{array}{cc}
                a m_{d} & b m_{s} V_{us} \\
                 & c m_{s}
        \end{array} \right) ,
\end{equation}
where $a$, $b$, $c$ are constants of order unity.  Unless $a=b=c$ 
exactly, $M_{d}$ and $m^{2,d}_{LR}$ are not simultaneously 
diagonalizable and we find
\begin{eqnarray}
        (\delta^{d}_{12})_{LR} \simeq \frac{m_{3/2} m_{s} 
        V_{us}}{m_{\tilde{q}}^{2}} = 2 \times 10^{-5}
        \left( \frac{m_{s}(M_{Pl})}{\rm 50~MeV}\right)
        \left(\frac{m_{3/2}}{m_{\tilde{q}}}\right)
        \left(\frac{\rm 500~GeV}{m_{\tilde{q}}}\right).
        \label{eq:estimate}
\end{eqnarray}
It turns out that, following the simplest implementation along
the lines of the above described idea, the amount of flavor changing LR
mass insertion in the s and d--squark propagator results to roughly
saturate the bound from $\varepsilon^\prime/\varepsilon$ if a SUSY phase
of order unity is present \cite{murayama}. 

This line of work has received a great deal of attention in recent times,
after the last experimental measurements of $\varepsilon^\prime/\varepsilon$
in KTeV and NA31 \cite{KTeV,NA31}. The effects of non--universal $A$ terms in
CP violation experiments were previously analyzed by Abel and Frere 
\cite{abel} and after this new measurement discussed in many different works
\cite{non-u}. In the following we show a complete realization of the above 
Masiero--Murayama (MM) mechanism from a Type I string--derived model
recently presented by one of the authors \cite{KKV}.
 
\subsection{Type I string model and $\varepsilon^\prime/\varepsilon$}
\label{sec:type1}

In first place we explain our starting model, which is based on type I 
string models. Our purpose is to study explicitly CP violation effects in 
models with non--universal gaugino masses and $A$--terms.
Type I models can realize such initial conditions.
These models contain nine--branes and three types of 
five--branes ($5_a$, $a=1,2,3$).
Here we assume that the gauge group $SU(3)\times U(1)_Y$ is on a 9--brane 
and the gauge group $SU(2)$ on the $5_1$--brane like in 
Ref.~\cite{brhlik2,ibrahim}, in order to get non--universal gaugino masses 
between $SU(3)$ and $SU(2)$.
We call these branes the $SU(3)$--brane and the $SU(2)$--brane, 
respectively.
\vskip 0.25cm

Chiral matter fields correspond to open strings spanning between 
branes. Thus, they must be assigned accordingly to their quantum numbers.
For example, the chiral field corresponding to the 
open string between the $SU(3)$ and $SU(2)$ branes has 
non--trivial representations under both $SU(3)$ and $SU(2)$, 
while the chiral field corresponding to the open string, 
which starts and ends on the $SU(3)$--brane, should be 
an $SU(2)$--singlet.
\vskip 0.25cm
There is only one type of the open string that spans between the 9 and 
5--branes, that we denote as the $C^{95_1}$.
However, there are three types of open strings which start and end on the 
9--brane, that is, the $C_i^9$ sectors (i=1,2,3), corresponding to the $i$--th
complex compact dimension among the three complex dimensions.
If we assign the three families to the different $C_i^9$ sectors 
we obtain non--universality in the right--handed sector.
Notice that, in this model, we can not derive non--universality for the 
squark doublets, i.e. the left--handed sector.
In particular, we assign the $C^{9}_1$ sector to the third family and
the $C^{9}_3$ and $C^{9}_2$, to the first and second families, respectively.
\vskip 0.25cm

Under the above assignment of the gauge multiplets and 
the matter fields, soft SUSY breaking terms are obtained,  
following the formulae in Ref.~\cite{typeI}.
The gaugino masses are obtained 
\begin{eqnarray}
\label{gaugino}
M_3 & = & M_1 = \sqrt 3 m_{3/2} \sin \theta  e^{-i\alpha_S}, \\
M_2 & = &  \sqrt 3 m_{3/2} \cos \theta \Theta_1 e^{-i\alpha_1}.
\end{eqnarray}
While the $A$--terms are obtained as 
\begin{equation}
A_{C_1^9}= -\sqrt 3 m_{3/2} \sin \theta e^{-i\alpha_S}=-M_3,
\label{A-C1}
\end{equation}
for the coupling including $C_1^{9}$, i.e. the third family, 
\begin{equation}
A_{C_2^9}= -\sqrt 3 m_{3/2}(\sin \theta e^{-i\alpha_S}+
\cos \theta (\Theta_1 e^{-i\alpha_1}- \Theta_2 e^{-i\alpha_2})),
\label{A-C2}
\end{equation}
for the coupling including $C_2^{9}$, i.e. the second 
family  and 
\begin{equation}
\label{A-C3}
A_{C_3^9}= -\sqrt 3 m_{3/2}(\sin \theta e^{-i\alpha_S}+
\cos \theta (\Theta_1 e^{-i\alpha_1}- \Theta_3 e^{-i\alpha_3})),
\end{equation}
for the coupling including $C_3^{9}$, i.e. the first family.
Here $m_{3/2}$ is the gravitino mass, $\alpha_S$ and $\alpha_i$ are 
the CP phases of the F--terms of the dilaton field $S$ and 
the three moduli fields $T_i$, and $\theta$ and $\Theta_i$ are 
goldstino angles, and we have the constraint, $\sum \Theta_i^2=1$.
\vskip 0.25cm 
Thus, if quark fields correspond to different 
$C_i^9$ sectors, we have non--universal A--terms.
We obtain the following A--matrix for both of the 
up and down sectors, 
\begin{eqnarray}
A= \left(
\begin{array}{ccc}
A_{C^9_3}  & A_{C^9_2} & A_{C^9_1} \\ A_{C^9_3} & A_{C^9_2} &
A_{C^9_1} \\ A_{C^9_3} & A_{C^9_2} & A_{C^9_1}
\end{array}
\right) \label{A-1}.
\end{eqnarray}
The trilinear SUSY breaking matrix, $(Y^A)_{ij}=(Y)_{ij}(A)_{ij}$, 
itself is obtained 
\begin{equation}
Y^A = \left(\begin{array}{ccc}
 &  &  \\  & Y_{ij} &  \\  &  & \end{array}
\right) \cdot 
\left(\begin{array}{ccc}
A_{C^9_3} & 0 & 0 \\ 0 & A_{C^9_2} & 0 \\ 0 & 0 & A_{C^9_1} \end{array}
\right),
\end{equation}
in matrix notation.
\vskip 0.25cm

In addition, soft scalar masses for quark doublets and 
the Higgs fields are obtained, 
\begin{equation}
\label{doublets}
m_{C^{95_1}}^2=m_{3/2}^2(1-{3 \over 2}\cos^2 \theta(1- 
\Theta_1^2)).
\end{equation}
The soft scalar masses for quark singlets are obtained as
\begin{equation}
\label{singlets}
m_{C_i^9}^2=m_{3/2}^2(1-3\cos^2 \theta \Theta^2_i),
\end{equation}
if it corresponds to  the $C_i^{9}$ sector.
\vskip 0.25cm

Now, below the string or SUSY breaking scale, this model is simply a MSSM 
with non--trivial soft--breaking terms
from the point of view of flavor. Scalar mass matrices and tri--linear terms
have completely new flavor structures, as opposed to the super--gravity 
inspired CMSSM or the SM, where the only connection 
between different generations is provided by the Yukawa matrices.

This model includes, in the quark sector, 7 different structures of flavor,
$M_{Q}^2$, $M_{U}^2$, $M_{D}^2$, $Y_d$, $Y_u$, $Y^A_d$ and $Y^A_u$. 
From these matrices, $M_{Q}^2$, the squark doublet mass matrix, is 
proportional to the identity matrix, and hence trivial, then we are left 
with 6 non--trivial flavor matrices.
Notice that we have always the freedom to diagonalize the hermitian
squark mass matrices (as we have done in the previous section, 
Eqs.(\ref{doublets},\ref{singlets})) and fix some general form for the
Yukawa and tri--linear matrices. In this case, these four matrices
are completely observable, unlike in the SM or CMSSM case.

At this point, to specify completely the model, we need not only the 
soft--breaking terms but also the complete Yukawa textures.
The only available experimental information is the Cabbibo--Kobayashi--Maskawa 
(CKM) mixing matrix and the quark masses. Here, we choose our Yukawa texture 
following two simple assumptions : i) the CKM mixing matrix originates from 
the down Yukawa couplings (as done in the MM case) and ii) our Yukawa 
matrices are hermitian \cite{RRR}.
With these two assumptions we fix completely the Yukawa matrices,
\begin{eqnarray}
\begin{array}{lr}
Y_u=\ \Frac{1}{v_2} \left(\begin{array}{ccc}
m_u & 0 & 0 \\ 0 & m_c & 0 \\ 0 & 0 & m_t \end{array}
\right)\ \  & \ \ \ \ 
Y_d=\ \Frac{1}{v_1}\  K^\dagger\ . \left(\begin{array}{ccc}
m_d & 0 & 0 \\ 0 & m_s & 0 \\ 0 & 0 & m_b \end{array}
\right) .\ K
\label{Yuk}
\end{array}
\end{eqnarray}
with $v= v_1 /(\cos \beta) = v_2 / (\sin \beta) = \sqrt{2} M_W / g$, and
$K$ the CKM matrix. We take $\tan \beta= v_2/v_1 = 2$ in the following in all
numerical examples. 
In this basis we can analyze the down tri--linear matrix,
\begin{equation}
Y^A_d (M_{St}) = \Frac{1}{v_1}\  K^\dagger\ .\ M_d\ .\ K\ .
\left(\begin{array}{ccc}
A_{C^9_3} & 0 & 0 \\ 0 & A_{C^9_2} & 0 \\ 0 & 0 & A_{C^9_1} \end{array}
\right)
\end{equation}
with $M_d=diag.(m_d, m_s, m_b)$.

Hence, together with the up tri--linear matrix we have our MSSM completely
defined. The next step is simply to use the MSSM Renormalization Group 
Equations \cite{RGE,BBM} to obtain the whole spectrum and couplings at the
low scale, $M_W$. The dominant effect in the tri--linear terms renormalization
is due to the gluino mass which produces the well--known alignment among
A--terms and gaugino phases. However, this renormalization is always 
proportional to the Yukawa couplings and not to the tri--linear terms, 
Eq.(\ref{Aurge}). This implies that, in the SCKM basis, the gluino effects 
will be diagonalized in excellent approximation, while due to the different
flavor structure of the tri--linear terms large off--diagonal elements will
remain with phases ${\cal{O}}(1)$ \cite{murayama}. To see this more explicitly, 
we can roughly approximate the RGE effects as,
\begin{equation}
Y^A_d (M_{W}) = c_{\tilde{g}}\  m_{\tilde{g}}\ Y_d \ +c_{A}\ Y_d\ .
\left(\begin{array}{ccc}
A_{C^9_3} & 0 & 0 \\ 0 & A_{C^9_2} & 0 \\ 0 & 0 & A_{C^9_1} \end{array}
\right)
\end{equation}
with $m_{\tilde{g}}$ the gluino mass and $c_{\tilde{g}}$, $c_A$ coefficients 
order 1 (typically $c_{\tilde{g}} \simeq 5$ and $c_A\simeq 1$).
\begin{figure}
\begin{center}
\epsfxsize = 13cm
\epsffile{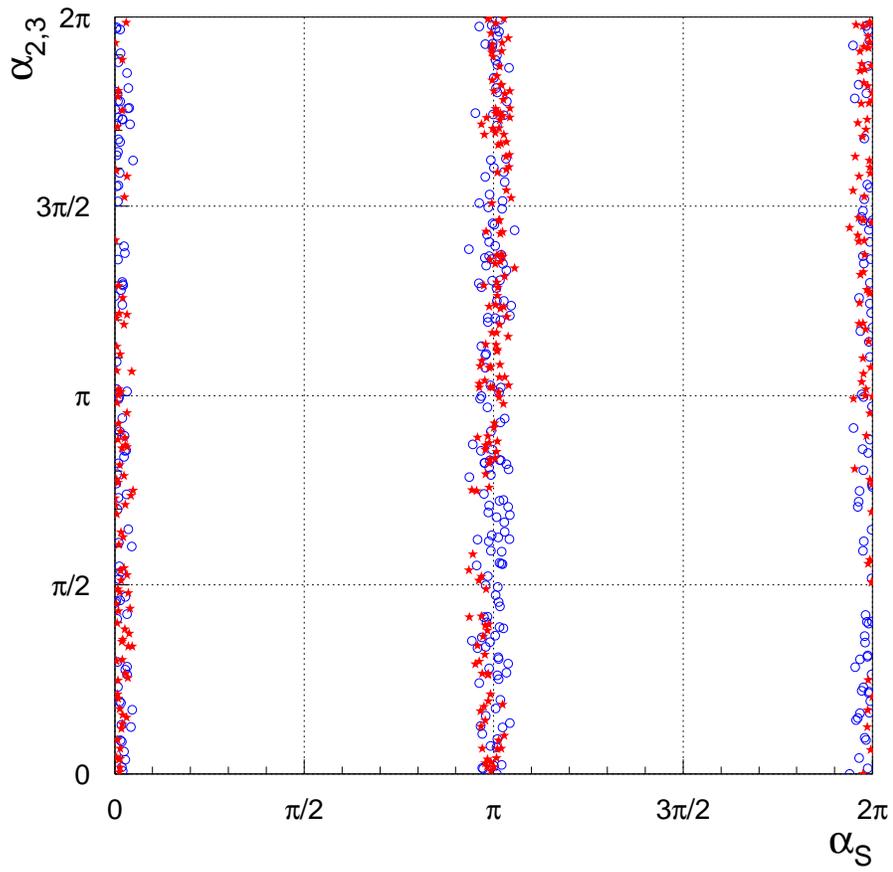}
\leavevmode
\end{center}
\caption{Allowed values for $\alpha_2$--$\alpha_S$ (open blue circles) and 
$\alpha_3$--$\alpha_S$ (red stars)}
\label{scat}
\end{figure}
We go to the SCKM basis after diagonalizing all the Yukawa matrices
(that is, $K . Y_d . K^\dagger = M_d /v_1$).
In this basis, we obtain the tri--linear couplings as,
\begin{equation}
v_1\ Y^A_d (M_{W}) = \Big(c_{\tilde{g}}\ m_{\tilde{g}}\ M_d \ +
c_{A}\ M_d\ .\ K\ .\left(\begin{array}{ccc}
A_{C^9_3} & 0 & 0 \\ 0 & A_{C^9_2} & 0 \\ 0 & 0 & A_{C^9_1} \end{array}
\right)\ .\ K^\dagger \Big)
\label{A-SCKM}
\end{equation}  
From this equation we can get the $L$--$R$ down squark mass matrix
\begin{equation} 
{m_{LR}^{2}}^{(d)}=v_1\ {Y^A_{d}}^* - \mu e^{i\varphi_{\mu}}
\tan\beta\, M_{d}
\end{equation}
And finally using unitarity of $K$ we obtain for the $L$--$R$ 
Mass Insertions,
\begin{eqnarray}
\label{Dlr}
(\delta_{LR}^{(d)})_{i j}= \frac{1}{m^2_{\tilde{q}}}\ m_i\ \Big(
\delta_{ij}\ (c_{A} A_{C^9_3}^*\ +\ c_{\tilde{g}}\ m_{\tilde{g}}^* -\ 
\mu e^{i\varphi_{\mu}} \tan\beta ) + \nonumber \\
K_{i 2}\ K^*_{j 2}\ c_{A}\ ( A_{C^9_2}^* - A_{C^9_3}^* ) +
K_{i 3}\ K^*_{j 3}\ c_{A}\ ( A_{C^9_1}^* - A_{C^9_3}^* ) \Big)
\end{eqnarray}
where $m^2_{\tilde{q}}$ is an average squark mass and $m_i$ the quark mass.
The same rotation must be applied to the $L$--$L$ and $R$--$R$ squark mass 
matrices,
\begin{eqnarray}
{M^{(d)}_{LL}}^2 (M_W)=   K\ .\ M_Q^2 (M_W)\ .\ K^\dagger \nonumber \\
{M^{(d)}_{RR}}^2 (M_W)=   K\ .\ M_D^2 (M_W)\ .\ K^\dagger 
\end{eqnarray}
However, the off--diagonal MI in these matrices are sufficiently small
in this case thanks to the universal and dominant contribution from gluino 
to the squark mass matrices in the RGE. 
\begin{figure}
\begin{center}
\epsfxsize = 13cm
\epsffile{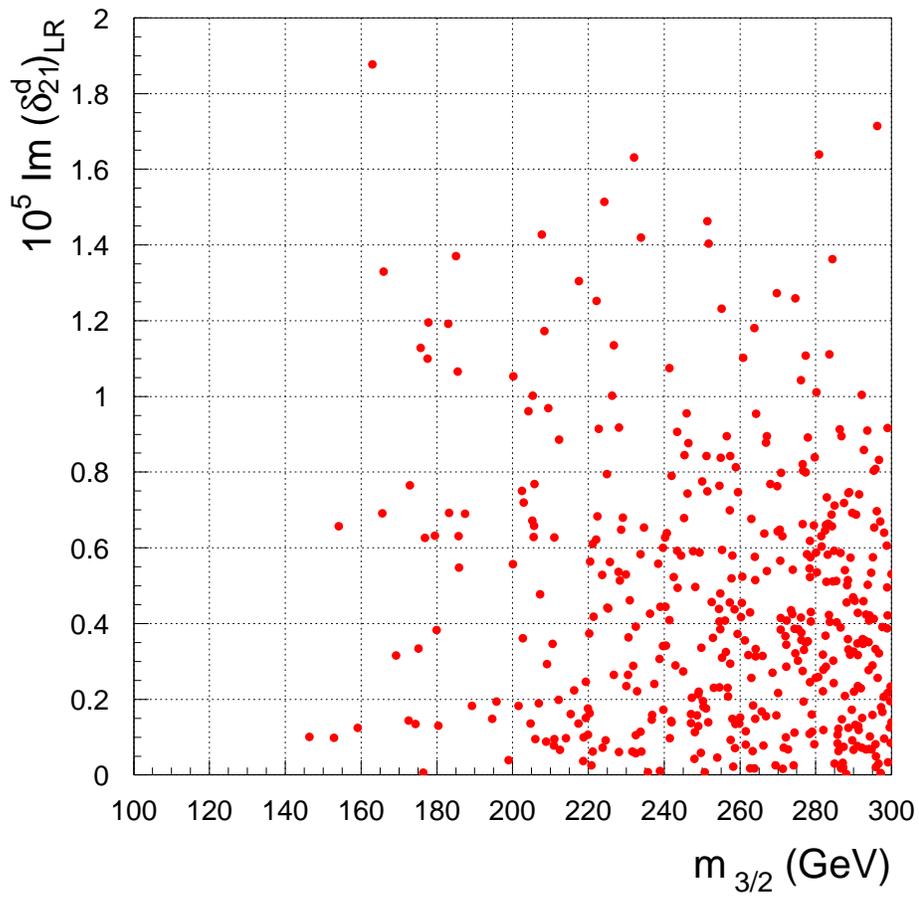}
\leavevmode
\end{center}
\caption{$(\delta_{LR}^{(d)})_{2 1}$ versus $m_{3/2}$ for experimentally 
allowed regions of the SUSY parameter space}
\label{eps'}
\end{figure}
At this point, with the explicit expressions for $(\delta_{LR}^{(d)})_{i j}$,
we can study the gluino mediated contributions to EDMs and 
$\varepsilon^\prime/\varepsilon$. In this non--universal scenario, it is
relatively easy to maintain the SUSY contributions to the EDM of the
electron and the neutron below the experimental bounds while having
large SUSY phases that contribute to $\varepsilon^\prime/\varepsilon$.
This is due to the fact the EDM are mainly controled by flavor--diagonal
MI, while gluino contributions to $\varepsilon^\prime/\varepsilon$ are
controled by $(\delta_{LR}^{(d)})_{1 2}$ and $(\delta_{LR}^{(d)})_{2 1}$.
Here, we can have a very small phase for $(\delta_{LR}^{(d)})_{1 1}$
and $(\delta_{LR}^{(u)})_{1 1}$ and phases ${\cal{O}}(1)$ for the 
off--diagonal elements without any fine--tuning \cite{KKV}. 
It is important to remember that the observable phase is always the relative 
phase between these mass insertions and the relevant gaugino mass involved.
In Eq.(\ref{Dlr}) we can see that the diagonal elements tend to align 
with the gluino phase, hence to have a small EDM, it is enough to have
the phases of the gauginos and the $\mu$ term approximately equal, 
$\alpha_S=\alpha_1= - \varphi_\mu$. However $\alpha_2$ and $\alpha_3$ can still
contribute to off--diagonal elements. In figure \ref{scat} we show
the allowed values for $\alpha_S$, $\alpha_2$ and $\alpha_3$ assuming
$\alpha_1=\varphi_\mu=0$. We impose the EDM, $\varepsilon_K$ and 
$b\rightarrow s \gamma$ bounds separately for gluino and chargino 
contributions together with the usual bounds on SUSY masses.
We can see that, similarly to the CMSSM situation, $\varphi_\mu$ is 
constrained to be very close to the gluino and chargino phases
(in the plot $\alpha_S \simeq 0, \pi$), but $\alpha_2$ and 
$\alpha_3$ are completely unconstrained.

Finally, in figure \ref{eps'}, we show the effects of these phases in the
$(\delta_{LR}^{(d)})_{2 1}$ MI as a function of the gravitino mass.
All the points in this plot satisfy all CP--conserving constraints besides
EDM and $\varepsilon_K$ constraints. We must remember that a value of 
$|{\rm Im}(\delta^{d}_{12})_{LR}^{2}| \sim 10^{-5}$ gives a significant 
contribution to $\varepsilon^\prime/\varepsilon$. In this plot, we can see
a large percentage of points above or close to $1 \times 10^{-5}$. 
Hence, we can conclude that, in the presence of new flavor structures in 
the SUSY soft--breaking terms, it is not difficult to obtain sizeable SUSY 
contributions to CP violation observables and specially to
$\varepsilon^\prime/\varepsilon$ \cite{murayama,KKV}.\footnote{With these 
$L$--$R$ mass insertions alone, it is in general difficult to saturate
$\varepsilon_K$ \cite{gabbiani}. However, in some special situations, 
it is still possible to have large contributions \cite{brhlik2,isidori}}

\section{Conclusions and Outlook}

Here we summarize the main points of these lectures:
\begin{itemize}
\item There exist strong theoretical and ``observational'' reasons to go
beyond the SM.
\item The gauge hierarchy and coupling unification problems favor the presence 
of low--energy SUSY (either in its minimal version, CMSSM, or more naturally,
in some less constrained realization).
\item Flavor and CP problems constrain low--energy SUSY, but, at the same
time, provide new tools to search for SUSY indirectly.
\item In all generality, we expect new CP violating phases in the SUSY
sector. However, these new phases are not going to produce sizeable
effects as long as the SUSY model we consider does not exhibit a new flavor 
structure in addition to the SM Yukawa matrices.
\item In the presence of a new flavor structure in SUSY, we showed that 
large contributions to CP violating observables are indeed possible.
\end{itemize}

In summary, given the fact that LEP searches for SUSY particles are 
close to their conclusion and that for Tevatron it may be rather challenging
to find a SUSY evidence, we consider CP violation a potentially precious 
ground for SUSY searches before the advent of the ``SUSY machine'', LHC. 
   
\section*{Acknowledgments}
We thank D. Demir, T. Kobayashi, S. Khalil and H. Murayama as co--authors of
some recent works reported in these lectures. We are grateful to 
S. Bertolini, L. Silvestrini and F.J. Botella for enlightening conversations. 
A.M. thanks the organizers for the stimulating settling in
which the school took place. The work of A.M. was partly supported by
the TMR project ``Beyond the Standard Model'' contract number ERBFMRX
CT96 0090; O.V. acknowledges financial support from a Marie Curie EC grant 
(TMR-ERBFMBI CT98 3087).


\begin{thebibliography}{99}
\bibitem{susy2}
J. Ellis and D.V. Nanopoulos, \plb{110}{82}{44};
\\
R. Barbieri and R. Gatto, \plb{110}{82}{211}.


\bibitem{susy1} 
For a phenomenologically oriented review, see:\\
P. Fayet and S. Ferrara, \prep{32C}{77}{249};
\\
H.P. Nilles, \prep{110C}{84}{1};
\\
H.E. Haber and G.L Kane, \prep{117C}{87}{1};
\\
For spontaneously broken N=1 supergravity, see:\\
E. Cremmer, S. Ferrara, L. Girardello and A. Van Proeyen, \npb{212}{83}{413}
and references therein;
\\
P. Nath, R. Arnowitt and A.H. Chamseddine, {\it Applied N=1 Supergravity} 
(World Scientific, Singapore, 1984);\\
A.G. Lahanas and D.V. Nanopoulos, \prep{145C}{87}{1}.


\bibitem{GMSB1}
M. Dine, W. Fischler and M. Srednicki, \npb{189}{81}{575};
\\
S. Dimopoulos and S. Raby, \npb{192}{81}{353};
\\
M. Dine and W. Fischler, \plb{110}{82}{227}; 
\\
M. Dine and M. Srednicki, \npb{202}{82}{238}; 
\\
M. Dine and W. Fischler, \npb{204}{82}{346};
\\
L. Alvarez--Gaum\'e, M. Claudson and M. Wise, \npb{207}{82}{96};
\\
C. Nappi and B. Ovrut, \plb{113}{82}{175}; 
\\
S. Dimopoulos and S. Raby, \npb{219}{83}{479}.


\bibitem{GMSB2}
A. Nelson and M. Dine, \prd{48}{93}{1277},
hep-ph/9303230;
\\
M. Dine, A. E. Nelson and Y. Shirman, \prd{51}{95}{1362},
hep-ph/9408384;
\\
M. Dine, A. Nelson, Y. Nir and Y. Shirman, \prd{53}{96}{2658},
hep-ph/9507378.


\bibitem{GMSB3}
E. Poppitz and S. Trivedi, \prd{55}{97}{5508},
hep-ph/9609529;
\\
N. Arkani--Hamed, J. March--Russel and H. Murayama, \npb{509}{98}{3}, 
hep-ph/9701286;
\\
H. Murayama, \prl{79}{97}{18},
hep-ph/9705271;
\\
S. Dimopoulos, G. Dvali, G. Giudice and R. Rattazzi, \npb{510}{98}{12}, 
hep-ph/9705307;
\\
S. Dimopoulos, G. Dvali and R. Rattazzi, \plb{413}{97}{336}, hep-ph/9707537
;\\
M. Luty, \plb{414}{97}{71}, hep-ph/9706554;
\\
T. Hotta, K.-I. Izawa and T. Yanagida, \prd{55}{97}{415},
hep-ph/9606203;
\\ 
N. Haba, N. Maru and T. Matsuoka, \npb{497}{97}{31},
hep-ph/9612468;
\\ 
L. Randall, \npb{495}{97}{37},
hep-ph/9612426;
\\
Y. Shadmi, \plb{405}{97}{99},
hep-ph/9703312;
\\
N. Haba, N. Maru and T. Matsuoka, \prd{56}{97}{4207},
hep-ph/9703250;
\\
C. Csaki, L. Randall and W. Skiba, \prd{57}{98}{383}, hep-ph/9707386;
\\
Y. Shirman, \plb{417}{98}{281}, hep-ph/9709383;
\\
G.F.~Giudice and R.~Rattazzi, 
hep-ph/9801271.


\bibitem{Dugan}
M. Dugan, B. Grinstein and L. Hall, \npb{255}{85}{413}.


\bibitem{Dimopoulos}
S. Dimopoulos and S. Thomas, \npb{465}{96}{23}, hep-ph/9510220.


\bibitem{EDMN}
W. Buchmuller and D. Wyler, \plb{121}{83}{321};
\\
J. Polchinski and M. Wise,  \plb{125}{83}{393};
\\
W. Fischler, S. Paban and S. Thomas, \plb{289}{92}{373},
hep-ph/9205233.


\bibitem{Ellis}
J. Ellis and R. Flores, \plb{377}{96}{83},
hep-ph/9602211.


\bibitem{Dine}
M. Dine, A. Nelson and Y. Shirman, in ref.~\cite{GMSB2};\\
M. Dine, Y. Nir and Y. Shirman, \prd{55}{97}{1501},
hep-ph/9607397.


\bibitem{CPSUSY}
M. J. Duncan and J. Trampetic, \plb{134}{84}{439};
\\
E. Franco and M. Mangano, \plb{135}{84}{445};
\\
J.M. Gerard, W. Grimus, A. Raychaudhuri and G. Zoupanos, 
\plb{140}{84}{349};
\\
J. M. Gerard, W. Grimus, A. Masiero, D. V. Nanopoulos and A. 
Raychaudhuri, \plb{141}{84}{79}; 
\npb{253}{85}{93};
\\
P. Langacker and R. Sathiapalan, \plb{144}{84}{401};
\\
M. Dugan, B. Grinstein and L. Hall, in ref.~\cite{Dugan}.


\bibitem{refbrignole}
A.~Brignole, F.~Feruglio and F.~Zwirner, \zpc{71}{96}{679},
hep-ph/9601293.


\bibitem{mpr}
M. Misiak, S. Pokorski and J. Rosiek, in {\it Heavy Flavours II}, eds. A.~Buras
and M.~Lindner (World Scientific), hep-ph/9703442.

\bibitem{branco}
G.C.~Branco, G.C.~Cho, Y.~Kizukuri and N.~Oshimo,
\npb{449}{95}{483};
\\
G.C.~Branco, G.C.~Cho, Y.~Kizukuri and N.~Oshimo,
\plb{337}{94}{316}, hep-ph/9408229.


\bibitem{cancel}
T. Ibrahim and P. Nath, \prd{58}{98}{111301}, hep-ph/9807501;
\\
M. Brhlik, G.J. Good and G.L. Kane, \prd{59}{99}{115004},
hep-ph/9810457;
\\
A.~Bartl, T.~Gajdosik, W.~Porod, P.~Stockinger and H.~Stremnitzer, 
\prd{60}{99}{073003},
hep-ph/9903402.

\bibitem{non-u}
S.A. Abel and J.M. Frere, \prd{55}{97}{1623},
hep-ph/9608251;
\\
S.~Khalil, T.~Kobayashi and A.~Masiero, \prd{60}{99}{075003}, 
hep-ph/9903544;
\\
S.~Khalil and T.~Kobayashi,
\plb{460}{99}{341}, hep-ph/9906374.


\bibitem{heavy}
S. Dimopoulos and G.F. Giudice, \plb{357}{95}{573},
hep-ph/9507282;
\\
A. Cohen, D.B. Kaplan and A.E. Nelson, \plb{388}{96}{599}, hep-ph/9607394;
\\
A. Pomarol and D. Tommasini, \npb{466}{96}{3}, hep-ph/9507462.

\bibitem{124}
Y.~Grossman, Y.~Nir and R.~Rattazzi, in {\it Heavy Flavours II}, eds. A.~Buras
and M.~Lindner (World Scientific), 
hep-ph/9701231.


\bibitem{flavor}
D. Demir, A. Masiero and O. Vives,  SISSA report n. SISSA/134/99/EP,
hep-ph/9911337.


\bibitem{CPcons}
D. Demir, A. Masiero and O. Vives,  \prdII{61}{00}{075009}, hep-ph/9909325.


\bibitem{typeI}
L.E.~Ibanez, C.~Munoz and S.~Rigolin, \npb{553}{99}{43},
hep-ph/9812397.


\bibitem{newcancel}
M.~Brhlik, L.~Everett, G.L.~Kane and J.~Lykken,
\prl{83}{99}{2124}, hep-ph/9905215;
\\
M.~Brhlik, L.~Everett, G.L.~Kane and J.~Lykken,
Fermilab preprint no. FERMILAB-PUB-99-230-T, August 1999, hep-ph/9908326;
\\
T.~Ibrahim and P.~Nath,
Santa Barbara U. preprint no. NSF-ITP-99-129, October 1999, hep-ph/9910553.


\bibitem{fully}
D.~Demir, A.~Masiero and O.~Vives,
\prl{82}{99}{2447}, Err. \ibid{83}{99}{2093}, hep-ph/9812337.


\bibitem{CPbs}
S.~Baek and P.~Ko,
\prl{83}{99}{488}, hep-ph/9812229;
\\
S.~Baek and P.~Ko, 
Korea Inst. Sci Report No. KAIST-TH-99-1, 1999, hep-ph/9904283.

\bibitem{gabbiani}
F.~Gabbiani, E.~Gabrielli, A.~Masiero and L.~Silvestrini,
\npb{477}{96}{321}, hep-ph/9604387.


\bibitem{BBM}
S.~Bertolini, F.~Borzumati, A.~Masiero and G.~Ridolfi,
\npb{353}{91}{591}.


\bibitem{RGE}
N.K.~Falck, \zpc{30}{86}{247}.

\bibitem{MI}
L.J.~Hall, V.A.~Kostelecky and S.~Raby, \npb{267}{86}{415}.


\bibitem{haber}
H.E. Haber and G.L Kane in ref.~\cite{susy1}.


\bibitem{cho}
P.~Cho, M.~Misiak and D.~Wyler,
\prd{54}{96}{3329},
hep-ph/9601360.

\bibitem{bsg}
F.M.~Borzumati,
\zpc{63}{94}{291}, hep-ph/9310212;
\\
S.~Bertolini and F.~Vissani,
\zpc{67}{95}{513}, hep-ph/9403397;
\\
T.~Goto, Y.Y.~Keum, T.~Nihei, Y.~Okada and Y.~Shimizu,
\plb{460}{99}{333}, hep-ph/9812369.

\bibitem{kagan-neubert}
A.L.~Kagan and M.~Neubert,
{\it Eur.\ Phys.\ J. } {\bf C7} (1999) 5,
hep-ph/9805303;
\\
A.L.~Kagan and M.~Neubert,
\prd{58}{98}{094012}, hep-ph/9803368.

\bibitem{ciuchini}
M.~Ciuchini {\it et al.}
\jhep{10}{98}{008}, hep-ph/9808328;
\\
R.~Contino and I.~Scimemi,
{\it Eur.\ Phys.\ J. }{\bf C10} (1999) 347,
hep-ph/9809437.

\bibitem{brhlik2}
M.~Brhlik, L.~Everett, G.L.~Kane, S.F.~King and O.~Lebedev,
Virginia Pol. Inst. Report no. VPI/IPPAP/99/08, Sept. 1999, hep-ph/9909480;

\bibitem{newflavor}
R.~Barbieri, R.~Contino and A.~Strumia, 
Pisa U. Report no. IFUP-TH-45-99, Aug. 1999, hep-ph/9908255.
\\
A.~L.~Kagan and M.~Neubert,
\prl{83}{99}{4429}, hep-ph/9908404.
\\
K.~S.~Babu, B.~Dutta and R.~N.~Mohapatra, Oklahoma State U. preprint no.
OSU-HEP-99-03,
hep-ph/9905464.

\bibitem{giudice}
E.~Gabrielli and G.~F.~Giudice, \npb{433}{95}{3},
hep-lat/9407029.

\bibitem{murayama}
A.~Masiero and H.~Murayama,
\prl{83}{99}{907}, hep-ph/9903363.

\bibitem{BIM} See, {\it e.g.}\/, Y.~Kawamura, H.~Murayama, and 
 M.~Yamaguchi, \prd{51}{95}{1337}, 
hep-ph/9406245;
\\
A.~Brignole, L.E.~Iba\~nez, and C.~Mu\~noz, in {\sl Perspectives on 
Supersymmetry}, {\it ed.}\/ G.~Kane, Singapore, World Scientific, 1998, 
hep-ph/9707209.

\bibitem{KTeV} 
KTeV Collaboration, A.~Alavi-Harati {\it et al.}, \prl{83}{99}{22},
hep-ex/9905060.

\bibitem{NA31} NA31 Collaboration (G.D.~Barr {\it et al.}\/), 
\plb{317}{93}{233}.

\bibitem{abel}
S.A. Abel and J.M. Frere in ref.~\cite{non-u}.

\bibitem{KKV}
S.~Khalil, T.~Kobayashi and O.~Vives, SISSA report number SISSA/23/20000/EP, 
hep-ph/0003086.

\bibitem{ibrahim}
T.~Ibrahim and P.~Nath in ref.~\cite{newcancel}


\bibitem{RRR}
P.~Ramond, R.~G.~Roberts and G.~G.~Ross, \npb{406}{93}{19},
hep-ph/9303320.

\bibitem{isidori}
G.~D'Ambrosio, G.~Isidori and G.~Martinelli, INFN preprint no. INFN-NA-99-39,
hep-ph/9911522.
\end{thebibliography}
\end{document}